\documentclass{article}

\usepackage{arxiv}

\usepackage[utf8]{inputenc} 
\usepackage[T1]{fontenc}    
\usepackage{xr-hyper}
\usepackage{hyperref}       
\usepackage{url}            
\usepackage{booktabs}       
\usepackage{amsfonts}       
\usepackage{nicefrac}       
\usepackage{microtype}      
\usepackage{amsmath}
\usepackage{amssymb}
\usepackage{graphicx}
\usepackage[table]{xcolor}

\definecolor{lightgray}{gray}{0.9}
\def\bSig\mathbf{\Sigma}

\title{A Hierarchical Integrative Group LASSO (HiGLASSO) framework for analyzing environmental mixtures}

\author{\textbf{Jonathan Boss$^{1}$, 
Alexander Rix$^{1}$,
Yin-Hsiu Chen$^{2}$,
Naveen N. Narisetty$^{3}$,
Zhenke Wu$^{1}$,} \\
\textbf{Kelly K. Ferguson$^{4}$,
Thomas F. McElrath$^{5}$,
John D. Meeker$^{6}$,
and 
Bhramar Mukherjee$^{1}$} \\ \\
$^{1}$Department of Biostatistics, University of Michigan, Ann Arbor, Michigan, U.S.A. \\
$^{2}$Google Inc., Mountain View, California, U.S.A. \\
$^{3}$Department of Statistics, University of Illinois at Urbana-Champaign, Champaign, Illinois, U.S.A. \\
$^{4}$Epidemiology Branch, National Institute of Environmental Health Sciences, Durham, North Carolina, U.S.A. \\
$^{5}$Department of Obstetrics and Gynecology, Brigham and Women's Hospital, Boston, Massachusetts, U.S.A. \\
$^{6}$Department of Environmental Health Sciences, University of Michigan, Ann Arbor, Michigan, U.S.A. \\ \\
Contact Information: bossjona@umich.edu, bhramar@umich.edu}

\begin{document}
\maketitle

\begin{abstract}
Environmental health studies are increasingly measuring multiple pollutants to characterize the joint health effects attributable to exposure mixtures. However, the underlying dose-response relationship between toxicants and health outcomes of interest may be highly nonlinear, with possible nonlinear interaction effects. Existing penalized regression methods that account for exposure interactions either cannot accommodate nonlinear interactions while maintaining strong heredity or are computationally unstable in applications with limited sample size. In this paper, we propose a general shrinkage and selection framework to identify noteworthy nonlinear main and interaction effects among a set of exposures. We design hierarchical integrative group LASSO (HiGLASSO) to (a) impose strong heredity constraints on two-way interaction effects (hierarchical), (b) incorporate adaptive weights without necessitating initial coefficient estimates (integrative), and (c) induce sparsity for variable selection while respecting group structure (group LASSO). We prove sparsistency of the proposed method and apply HiGLASSO to an environmental toxicants dataset from the LIFECODES birth cohort, where the investigators are interested in understanding the joint effects of 21 urinary toxicant biomarkers on urinary 8-isoprostane, a measure of oxidative stress. An implementation of HiGLASSO is available in the \textsf{higlasso} R package, accessible through the Comprehensive R Archive Network.
\end{abstract}


\section{Introduction}
\label{s:intro}

Studying the effects of chemical exposures and their interactions in relation to adverse health outcomes is an important topic in epidemiological and environmental research. Furthermore, exposure to endocrine disruptors, such as phthalates and phenols, is of particular interest due to the ubiquity of exposure in the U.S. general population \cite{crinnion2010}. Phthalates are a group of chemicals that are widely used as plasticizers or solvents in products such as food packaging, cosmetics, and other industrial materials, which typically enter the human body through daily ingestion and inhalation \cite{schettler2006}. Phthalates are known for anti-androgenic effects and reproductive toxicity and recent studies have reported that the modes of their action include mechanisms such as oxidative stress \cite{ferguson2011, ferguson2012}. Phenols are a class of chemical compounds used in the manufacture of polycarbonate plastics and epoxy resins. Applications of some phenols include use in pesticides and personal care products such as makeups and toothpastes \cite{darbre2008}. Phenols may possess estrogenic activity and are linked to higher levels of maternal oxidative stress, inflammation in pregnancy, and reduced fetal growth \cite{watkins2015,ferguson2018}.

Classical environmental epidemiology has focused on analyzing one toxicant at a time even though, in truth, subjects are simultaneously exposed to a mixture of compounds which may work in concert. Namely, potential synergistic and antagonistic effects of chemical mixtures have been minimally addressed in human studies. The primary reasons behind single toxicant analysis are the lack of studies with measures on multiple pollutants and a lack of a principled analytic strategy for understanding effects of multiple pollutants and their interactions with limited sample size. Modern assaying technology has made it possible to measure multiple pollutants on the same subjects and advances in statistical learning have enabled us to develop methods that capture nonlinearity and interactions in complex exposure-response surfaces. Commonly used approaches that characterize the joint effects of mixtures on health outcomes in a flexible way include classification and regression tree (CART) \cite{loh2011} and Bayesian kernel machine regression (BKMR) \cite{bobb2015}. However, the number of candidate effects, including main effects and interaction effects, may be much larger than the number of observations (i.e., $p>>n$). To address this issue, one common approach is to introduce sparsity during estimation to shrink coefficient estimates towards a subset of variables that have stronger effects. This paper proposes a variable selection framework to handle potential nonlinearity and interactions between a set of multiple exposures. We then apply this framework to data from the LIFECODES study, an ongoing prospective pregnancy/birth cohort at Brigham and Women's Hospital (BWH), to identify important exposures and two-way exposure interactions that are associated with 8-isoprostane, an oxidative stress biomarker \cite{ferguson2015a}.

There are two major classes of methods for variable selection: penalty-based methods and forward/stepwise selection methods. The former adds a penalty term to an objective function which, upon optimization, induces sparsity. Some examples include the $L_1$ penalty in LASSO \cite{tibshirani1996}, the $L_0$ penalty in nonnegative garrote \cite{breiman1995}, the $L_{\gamma}$ penalty with $\gamma \geq 1$ in bridge regression \cite{fu1998}, the mixture of $L_1$ and $L_2$ penalties in elastic-net \cite{zou2005}, and the smoothly clipped absolute deviation (SCAD) penalty \cite{fan2001}. These methods can be used to incorporate interactions by treating interaction terms as additional predictors. However, including interaction terms in the absence of at least one corresponding main effect deviates from a naturally interpretable hierarchical interaction structure. \cite{nelder1977} and \cite{mccullagh1989} introduced the concepts of weak/strong heredity and marginality respectively as conceptual constraints to simplify model interpretation \cite{mccullagh1984} and improve statistical power \cite{cox1984}. Recent penalty-based methods that respect these heredity principles include the strong heredity interaction model (SHIM) \cite{choi2010}, the LASSO for hierarchical interactions (hierNet) \cite{bien2013}, and the group-LASSO interaction network (GLinternet) \cite{lim2015}. In addition to penalization based methods, forward selection methods \cite{boos2009, wasserman2009, luo2015} are also commonly used for variable selection in practice. Several forward selection methods which incorporate heredity constraints with linear and nonlinear interactions have been proposed \cite{wu2010, crews2011, hao2014, narisetty2018}.

Nonlinear exposure-response relationships have also been explored in environmental studies. Failure to account for nonlinearity could result in important variables being left out. Moreover, not properly adjusting for nonlinear main effects might result in spurious detection of interaction effects \cite{cornelis2012, mukherjee2012, he2017}. For example, the quadratic main effect terms of two predictors and interaction terms between the two predictors are not easily differentiable in practice, especially when the signal-to-noise level is low \cite{he2017}. Group LASSO \cite{yuan2006} can be adopted to model nonlinear effects where each group of variables represents the nonlinear expansion of a single predictor with respect to a chosen basis \cite{huang2010}. Another work that considers modeling nonlinear main effects using penalization is the COmponent Selection and Smoothing Operator (COSSO) \cite{lin2006}. To our knowledge, Variable selection using Adaptive Nonlinear Interaction Structures in High dimensions (VANISH) \cite{radchenko2010} is the only existing method that accounts for both nonlinear main and interaction effects with strong heredity enforced.

Using the same tuning parameter $\lambda$ (degree of penalization) for each predictor/group without assessing their relative importance may simultaneously reduce estimation efficiency and affect selection consistency \cite{leng2006}. Adaptive shrinkage has been extensively discussed in previous literature \cite{wang2007, zhang2007}. For example, adaptive LASSO \cite{zou2006}, adaptive elastic-net \cite{zou2009}, and adaptive Group LASSO \cite{wang2008} assign a separate penalty to each predictor/group, usually determined by the reciprocal of the absolute values of the corresponding coefficients. This ensures that smaller coefficients are shrunk to zero faster whereas larger coefficients are less penalized. Ordinary least squares (OLS) can be used to estimate the coefficients, however, when $p>n$, OLS cannot be implemented. In addition, it could be difficult for an analyst to supply a $\sqrt{n}$-consistent estimate of main and interaction effects to use as adaptive weights in a high-dimensional scenario, in which case oracle properties are not maintained. In this paper, we bypass the need to specify a set of initial coefficient estimates by using integrative weighted group LASSO \cite{pan2016} which jointly estimates weights and coefficients. 

We propose \textit{hierarchical integrative group LASSO (HiGLASSO)}, to deal with both nonlinear main and interaction effects under strong heredity while incorporating integrative weights for improved selection properties. The rest of the article is organized as follows. We briefly review the existing penalty-based interaction selection methods with heredity constraints in Section \ref{existing}. In Section \ref{HiGLASSO}, we describe HiGLASSO, the optimization procedure, and prove sparsistency for the resulting HiGLASSO estimator. We examine the performance of HiGLASSO by comparing it to other procedures that address nonlinearity, interaction terms, and/or group structure in Section \ref{simulation}. In Section \ref{BWH}, we analyze data from the LIFECODES study to identify important phthalates, phenols, and their possible interactions that associate with the oxidative stress biomarker 8-isoprostane. We conclude with a discussion in Section \ref{discussion}.

\section{Review of existing penalty-based interaction selection methods with heredity constraints}
\label{existing}
First we overview existing penalized regression methods that select interaction terms subject to heredity constraints. Consider the standard regression setting with $p$ predictors and $n$ observations where $\boldsymbol{x}_j$ denotes the $n \times 1$ predictor vector corresponding to the $j^{th}$ regression coefficient $\beta_j$, for $j = 1, \cdots, p$. Let $\gamma_{kl}$ be the coefficient of interaction effect between $\boldsymbol{x}_k$ and $\boldsymbol{x}_l$. Strong and weak heredity principles for interaction effects are defined as follows.
\begin{itemize}
\item \underline{Strong heredity principle:} If an interaction term is included in the model \textit{both} of its corresponding main effects must be present in the model. That is, if $\gamma_{kl} \neq 0$, then $\beta_k \neq 0$ \textit{and} $\beta_l \neq 0$.
\item \underline{Weak heredity principle:} If an interaction term is included in the model,\textit{ at least one} of the corresponding main effects must be present in the model. That is, if $\gamma_{kl} \neq 0$, then $\beta_k \neq 0$ \textit{or} $\beta_l \neq 0$.
\end{itemize}

\subsection{Methods for linear interactions}
A generic second-order model accounting for pairwise interaction effects with linear predictors is given as
\begin{equation}
\label{linint}
    \boldsymbol{y}=\boldsymbol{X}\boldsymbol{\beta}+\boldsymbol{X}_{(I)} \hspace{0.5 mm} \boldsymbol{\gamma} + \boldsymbol{\epsilon}
\end{equation}
where $\boldsymbol{X}=[\boldsymbol{x}_1, \cdots, \boldsymbol{x}_p]$ denotes the $n \times p$ design matrix for main effects, $\boldsymbol{X}_{(I)} = [\boldsymbol{x}_1 \odot \boldsymbol{x}_2, \cdots, \boldsymbol{x}_{p-1} \odot \boldsymbol{x}_p]$ denotes the $n \times [p(p-1)/2]$ design matrix for interactions where $``\odot"$ indicates the element-wise product, $\boldsymbol{\beta}=(\beta_1, \cdots, \beta_p)^{\top}$, $\boldsymbol{\gamma} = (\gamma_{12}, \cdots, \gamma_{p-1, p})^{\top}$, and $\boldsymbol{\epsilon}$ is a multivariate Gaussian error vector. Without loss of generality, we assume all variables are standardized and exclude the intercept from our regression model. We first review existing methods for selecting interaction effects which satisfy the strong heredity principle.

\textbf{SHIM (Strong heredity interaction model)} \cite{choi2010}: SHIM reparametrizes the interaction coefficients as scaled products of component main effect terms, namely $\gamma_{ij}=\eta_{ij}\beta_i\beta_j$ for $1 \leq i < j \leq p$, $\eta_{ij} \in \mathbb{R}$. A penalty is imposed on $\boldsymbol{\eta} = \{\eta_{ij}\}$ rather than the interaction coefficients $\boldsymbol{\gamma}$ to preserve heredity of the interaction terms in the selected model. SHIM minimizes the objective function
\begin{equation*}
 \frac{1}{2}||\boldsymbol{y}-\boldsymbol{X\beta}-\boldsymbol{X}_{(I)}\boldsymbol{\gamma}||^2_2+\lambda_1 ||\boldsymbol{\beta}||_1 + \lambda_2 ||\boldsymbol{\eta}||_1
\end{equation*}
using an algorithm that iterates between LASSO and group LASSO.

\textbf{hierNet} \cite{bien2013}: hierNet is a LASSO-based approach which minimizes
\begin{equation*}
 \frac{1}{2}\Big\Vert\boldsymbol{y}-\boldsymbol{X\beta}-\sum_{k=1}^{p}\sum_{l=1}^p(\boldsymbol{x}_k \cdot \boldsymbol{x}_l)\gamma_{kl}\Big\Vert_2^2+\lambda_1 \sum_{j=1}^{p}|\beta_j| + \frac{1}{2}\lambda_2 \sum_{k=1}^{p} \sum_{l=1}^p |\gamma_{kl}|,
\end{equation*}
subject to symmetry constraints $\gamma_{kl}=\gamma_{lk}$, $\forall$ $1 \leq k, l \leq p$, and heredity constraints $\sum_{l=1}^p |\gamma_{kl}| \leq |\beta_{k}|$, $\forall k = 1, \cdots, p$, which ensure that the interaction effects are zero given that any of the corresponding main effects are zero. Alternating Direction Method of Multipliers (ADMM) \cite{boyd2011} is used to solve the constrained optimization.

\textbf{GLinternet (group-LASSO interaction network)} \cite{lim2015}: GLinternet uses an overlapping group LASSO penalty to enforce strong heredity. The objective function is given by
\begin{equation*}
 \frac{1}{2}\Big\Vert\boldsymbol{y}-\boldsymbol{X\beta}-\sum_{k=2}^{p}\sum_{l=1}^{k-1}[\boldsymbol{x}_k , \boldsymbol{x}_l, \boldsymbol{x}_k \cdot \boldsymbol{x}_l]\boldsymbol{\gamma}_{kl}^*\Big\Vert_2^2+\lambda_1 ||\boldsymbol{\beta}||_1 + \frac{1}{2}\lambda_2 \sum_{k=2}^{p} \sum_{l=1}^{k-1} ||\boldsymbol{\gamma}_{kl}^*||_2,
\end{equation*}
where each $\boldsymbol{\gamma}_{kl}^*$ is a three dimensional vector with the first two elements corresponding to main effects and the third element corresponding to the interaction effect. Note that the main effects appear multiple times inside the $L_2$-norm (parameterized by $\beta_k$, $\boldsymbol{\gamma}_{kl}^*$ for $l < k$, and $\boldsymbol{\gamma}_{lk}^*$ for $l > k$) and hence are multiply penalized. An iterative soft thresholding algorithm \cite{beck2009} can be used to solve the GLinternet optimization problem.

\subsection{Methods for nonlinear interactions}
\label{existingnonlin}
Basis functions such as cubic splines are often used to incorporate nonlinear main effects and nonlinear interaction effects into regression models. Consider $S$ groups of predictors each of which corresponds to a pre-specified nonlinear basis expansion. Let $\boldsymbol{X}_j$ and $\boldsymbol{\beta}_j$ denote the $n \times p_j$ design matrix and coefficient vector of length $p_j$ corresponding to group $j$ of basis size $p_j$, respectively, for $j = 1, \cdots, S$. Let $\boldsymbol{X}_{kl}$ be the $n \times (p_kp_l)$ design matrix for two-way interaction between group $k$ and group $l$ and $\boldsymbol{\gamma}_{kl}$ be the corresponding $(p_kp_l)-$vector of interaction coefficients for $1 \leq k < l \leq S$. Note that $\boldsymbol{X}_j$ and $\boldsymbol{X}_{kl}$ are distinct from their section 2.1 counterparts, $\boldsymbol{X}$ and $\boldsymbol{X}_{(I)}$, because we are now working with basis expansions of exposures rather than linear exposure terms. We focus on the second-order model with interaction effects for $S$ groups of nonlinear predictors as
\begin{equation}
\label{nonlinint}
    \boldsymbol{y}=\sum_{j=1}^S\boldsymbol{X}_j\boldsymbol{\beta}_j+\sum_{1 \leq k < l \leq S}\boldsymbol{X}_{kl}\boldsymbol{\gamma}_{kl} + \boldsymbol{\epsilon}.
\end{equation}
VANISH is the only existing penalty-based method that imposes sparsity and strong heredity on model (\ref{nonlinint}).

\textbf{VANISH} \cite{radchenko2010}: VANISH optimizes penalized least squares as
\begin{align*}
 & \frac{1}{2}\Big\Vert\boldsymbol{y}-\sum_{j = 1}^S\boldsymbol{X}_j\boldsymbol{\beta}_j-\sum_{1 \leq k < l \leq S}\boldsymbol{X}_{kl}\boldsymbol{\gamma}_{kl}\Big\Vert_2^2 +\lambda_1 \sum_{j=1}^{S}\Big(||\boldsymbol{\beta}_j||_2^2+\sum_{k<j}||\boldsymbol{\gamma}_{kj}||_2^2 +\sum_{l>j}||\boldsymbol{\gamma}_{jl}||_2^2\Big)^{1/2} + \lambda_2 \sum_{1 \leq k < l \leq S}||\boldsymbol{\gamma}_{kl}||_2. &
\end{align*}
By construction, $\boldsymbol{\beta}_j$'s and $\boldsymbol{\gamma}_{kl}$'s are folded together in the first penalty term so main effect coefficients and interaction coefficients are either all zero or all nonzero, based on the property of the group LASSO penalty. The same structure applies to all $S$ groups of main effects so strong heredity is guaranteed. A block gradient descent algorithm involving a single sweep through all the variables is applied to obtain a solution to the VANISH objective function.

None of the existing variable selection methods described so far account for both strong heredity in interaction selection and differential penalization via adaptive weighting. We propose HiGLASSO as a novel approach to select two-way interaction effects under strong heredity constraints using penalization with integrative weights, circumventing the need for initial coefficient estimates.

\section{Hierarchical integrative group LASSO (HiGLASSO)}
\label{HiGLASSO}
\subsection{HiGLASSO formulation}
Consider the regression model in (\ref{nonlinint}). To enforce heredity constraints, we rewrite (\ref{nonlinint}) as
\begin{equation}
\label{nonlinint2}
    \boldsymbol{y} =\sum_{j=1}^{S}\boldsymbol{X}_j\boldsymbol{\beta}_j+\sum_{1 \leq j<j' \leq S}\boldsymbol{X}_{jj'}[\boldsymbol{\eta}_{jj'} \odot (\boldsymbol{\beta}_j \otimes \boldsymbol{\beta}_{j'})]+\boldsymbol{\epsilon}
\end{equation}
by reparameterizing $\boldsymbol{\gamma}_{jj'}=\boldsymbol{\eta}_{jj'} \odot (\boldsymbol{\beta}_j \otimes \boldsymbol{\beta}_{j'})$ for $1 \leq j < j' \leq S$. Here $``\otimes"$ denotes the Kronecker product and $\boldsymbol{\eta}_{jj'}$ is a $(p_jp_{j'})-$vector of scalars for interactions between variables in group $j$ and group $j'$ following SHIM \cite{choi2010}. Note that strong heredity constraints are satisfied because $\boldsymbol{\gamma}_{jj'}$ is non-zero only if both main effects are non-zero. To see this, $\boldsymbol{\beta}_j=\boldsymbol{0}$ and/or $\boldsymbol{\beta}_{j'}=\boldsymbol{0}$ implies that $\boldsymbol{\gamma}_{jj'}=\boldsymbol{0}$. Similarly, $\boldsymbol{\gamma}_{jj'} \neq \boldsymbol{0}$ implies that $\boldsymbol{\eta}_{jj'} \neq \boldsymbol{0}$, $\boldsymbol{\beta}_j \neq \boldsymbol{0}$, and $\boldsymbol{\beta}_{j'} \neq \boldsymbol{0}$.

Consider the penalized least squares criterion
\begin{align}
  & \underset{\boldsymbol{\beta}_j, \boldsymbol{\eta}_{jj'}}{\arg\min} \hspace{1 mm} \frac{1}{2}\Big\Vert\boldsymbol{y}-\sum_{j=1}^{S}\boldsymbol{X}_j\boldsymbol{\beta}_j-\sum_{1 \leq j<j' \leq S}\boldsymbol{X}_{jj'}[\boldsymbol{\eta}_{jj'} \odot (\boldsymbol{\beta}_j \otimes \boldsymbol{\beta}_{j'})]\Big\Vert_2^2 &  \label{HiGLASSOcriterion0} \\
  & + \lambda_1\sum_{j = 1}^{S}||\boldsymbol{\beta}_j||_2 + \lambda_2\sum_{1 \leq j<j' \leq S}||\boldsymbol{\eta}_{jj'}||_2, \nonumber &
\end{align}
where $\lambda_1$ and $\lambda_2$ are tuning parameters that control the amount of main effect and interaction effect shrinkage toward 0, respectively. To remedy potential estimation inefficiency and selection inconsistency, we work with a modified version of (\ref{HiGLASSOcriterion0}) to differentially penalize parameters in the spirit of adaptive group LASSO \cite{wang2008}. We consider
\begin{align}
 & \underset{\boldsymbol{\beta}_j, \boldsymbol{\eta}_{jj'}}{\arg\min} \hspace{1 mm} \frac{1}{2}\Big\Vert\boldsymbol{y}-\sum_{j=1}^{S}\boldsymbol{X}_j\boldsymbol{\beta}_j-\sum_{1 \leq j<j' \leq S}\boldsymbol{X}_{jj'}[\boldsymbol{\eta}_{jj'} \odot (\boldsymbol{\beta}_j \otimes \boldsymbol{\beta}_{j'})]\Big\Vert_2^2 & \label{HiGLASSOcriterion} \\
  & + \lambda_1\sum_{j = 1}^{S}w_j||\boldsymbol{\beta}_j||_2 + \lambda_2\sum_{1 \leq j<j' \leq S}w_{jj'}||\boldsymbol{\eta}_{jj'}||_2, \nonumber &
\end{align}
where $w_j$'s and $w_{jj'}$'s are pre-specified weight functions of unknown coefficients $\{ \boldsymbol{\beta}_j \}$ and $\{ \boldsymbol{\eta}_{jj'} \}$.

\begin{figure}[!htbp]
\centering
\scalebox{0.8}{\includegraphics[width=1.0\columnwidth]{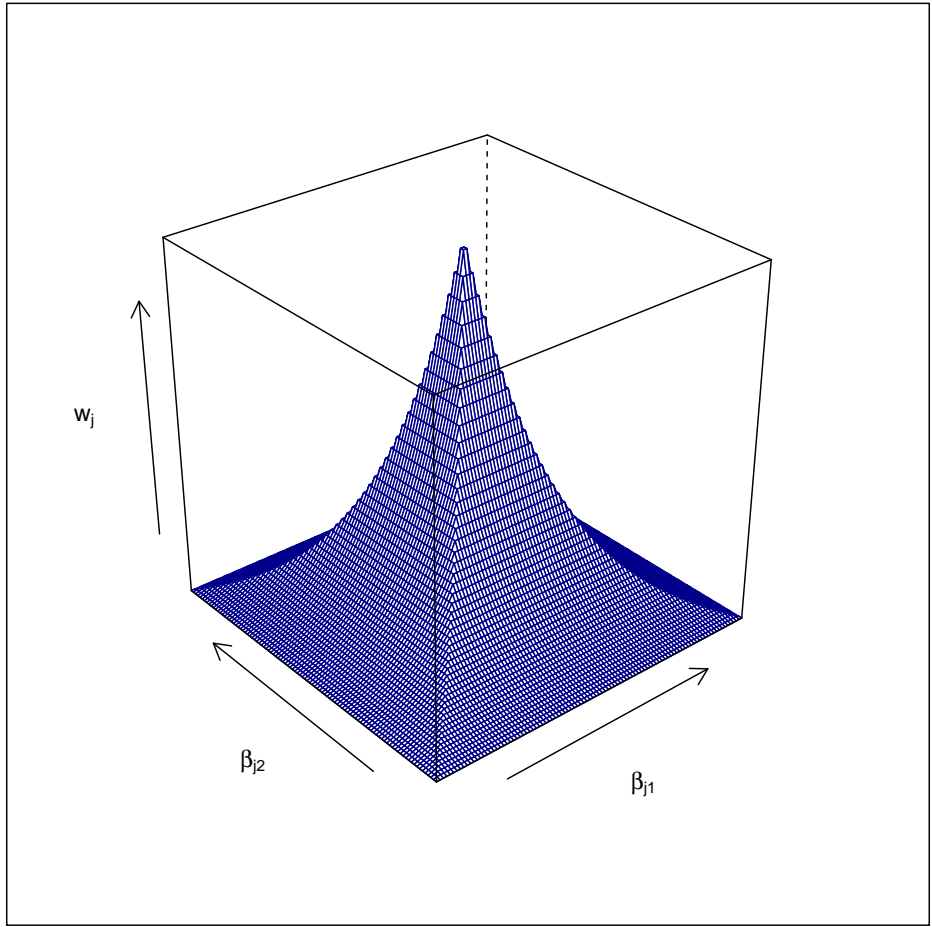}}
\caption{HiGLASSO weight function evaluated for a two-dimensional vector in $[-3, 3] \times [-3, 3]$ with $\sigma = 1$.}
\label{weight_function}
\end{figure}

To concurrently estimate weights and model parameters following \cite{pan2016}, we consider weight functions based on the extreme values of each group, namely,
\begin{equation}
\label{weightfn1}
    w_j \equiv \text{exp}\bigg\{-\frac{||\boldsymbol{\beta}_j||_{\infty}}{\sigma}\bigg\} \text{ for } j = 1, \cdots, S,
\end{equation}
\begin{equation}
\label{weightfn2}
w_{jj'} \equiv \text{exp}\bigg\{-\frac{||\boldsymbol{\eta}_{jj'}||_{\infty}}{\sigma}\bigg\} \text{ for } 1 \leq j < j' \leq S,
\end{equation}
where $||\boldsymbol{\mu}||_{\infty}$ is the $L_{\infty}$ norm of $\boldsymbol{\mu}$ and $\sigma$ is a pre-determined scale parameter. That is, the weights decay exponentially with the extremum norm of the coefficients within a group. Figure \ref{weight_function} illustrates the weight function for a two-dimensional coefficient vector. We adopt the $L_{\infty}$ norm, instead of the $L_0$, $L_1$, and $L_2$ norms, because the groups in our motivating example are basis expansions of each exposure. We do not want to impose sparsity within each group; therefore, to assess the effect size of the entire basis expansion, taking the extremum of the coefficients within a group is more meaningful than taking an ``average" coefficient.

In summary, HiGLASSO has the following four features:
\begin{enumerate}
\item Imposes strong heredity on two-way interaction (Hierarchical);
\item Incorporates adaptive weights without requiring initial coefficient estimates (Integrative);
\item Induces sparsity for variable selection (LASSO);
\item Maintains group structure (Group LASSO). The HiGLASSO framework is general and the group structure can be defined based on the specific application. For example, the group structure could be:
\begin{itemize}
\item A set of basis functions representing nonlinear relationships,
\item Dummy variables representing different levels of categorical variables,
\item A natural grouping based on domain knowledge.
\end{itemize}
\end{enumerate}

\subsection{Optimizing the HiGLASSO objective function}
The objective function in \eqref{HiGLASSOcriterion} is non-convex and is difficult to globally minimize, however \cite{pan2016} proposed a generalized local quadratic approximation which we utilize to find a local minimum. The first term in \eqref{HiGLASSOcriterion} involves the product of $\boldsymbol{\beta}_j$'s and $\boldsymbol{\eta}_{jj'}$'s. We use an iterative approach to cycle through $\boldsymbol{\beta}_1, \cdots, \boldsymbol{\beta}_S$, and the $\boldsymbol{\eta}_{jj'}$'s until convergence using gradient descent. We first optimize over $\boldsymbol{\beta}_j$ given the current $\hat{\boldsymbol{\beta}}_{j'}$'s with $j' \neq j$ and $\hat{\boldsymbol{\eta}}_{jj'}$'s. Then we iteratively obtain $\hat{\boldsymbol{\eta}}_{jj'}$ estimates given current $\hat{\boldsymbol{\beta}}_j$'s. The optimization routine is summarized in Web Appendix \textcolor{cyan}{A}. The \textsf{higlasso} R package, available on the Comprehensive R Archive Network (CRAN), implements the proposed optimization routine.

\subsection{Sparsistency of HiGLASSO estimator}
\label{estimation}
We now establish sparsistency of the HiGLASSO estimator obtained as the minimizer of \eqref{HiGLASSOcriterion}. Let $\boldsymbol{\theta}$ denote the vector of all coefficients, including main effect coefficients and interaction coefficients. Namely, $\boldsymbol{\theta} = (\boldsymbol{\beta}^{\top}, \boldsymbol{\gamma}^{\top})^{\top}$ where $\boldsymbol{\beta}=(\boldsymbol{\beta}_1^{\top}, \cdots, \boldsymbol{\beta}_S^{\top})^{\top}$, $\boldsymbol{\gamma}=(\boldsymbol{\gamma}_{12}^{\top}, \cdots, \boldsymbol{\gamma}_{S-1, S}^{\top})^{\top}$, and $\boldsymbol{\gamma}_{jj'} = \boldsymbol{\eta}_{jj'} \odot (\boldsymbol{\beta}_j \otimes \boldsymbol{\beta}_{j'})$. Denote $\boldsymbol{\theta}_{\mathcal{P}}=(\boldsymbol{\beta}_{\mathcal{P}_1}{}^{\top}, \boldsymbol{\gamma}_{\mathcal{P}_2}{}^{\top})^{\top}$ and $\boldsymbol{\theta}_{\mathcal{P}^{\mathsf{c}}}=(\boldsymbol{\beta}_{\mathcal{P}_1^{\mathsf{c}}}{}^{\top}, \boldsymbol{\gamma}_{\mathcal{P}_2^{\mathsf{c}}}{}^{\top})^{\top}$ where ${\mathcal{P}_1}$ is the true nonzero set for $\boldsymbol{\beta}$, $\mathcal{P}_1^{\mathsf{c}}$ is the true zero set for $\boldsymbol{\beta}$, ${\mathcal{P}_2}$ is the true nonzero set for $\boldsymbol{\gamma}$, $\mathcal{P}_2^{\mathsf{c}}$ is the true zero set for $\boldsymbol{\gamma}$, $\mathcal{P} = \mathcal{P}_1 \cup \mathcal{P}_2$, and $\mathcal{P}^{\mathsf{c}} = \mathcal{P}_1^{\mathsf{c}} \cup \mathcal{P}_2^{\mathsf{c}}$. Let $a_n = \min (\lambda_1(n), \lambda_2(n))$ and $b_n = \sigma(n)$. That is, $\lambda_1(n)$, $\lambda_2(n)$, and $\sigma(n)$ depend on sample size. \\

\noindent \textbf{Theorem} (Sparsistency of HiGLASSO estimator): Suppose that the data are generated from the model given by \eqref{nonlinint2} with the errors $\epsilon$ following an i.i.d. normal distribution with mean zero and variance $\tau^2 >0$. Assume that the design matrix $\boldsymbol{X}$ is random such that $\frac{1}{n}\boldsymbol{X}^{\intercal}\boldsymbol{X} = \frac{1}{n}E \left( \boldsymbol{X}^{\intercal}\boldsymbol{X} \right) + O_p(n^{-1/2}),$  $\frac{1}{n}E \left( \boldsymbol{X}^{\intercal}\boldsymbol{X} \right)$ is invertible, all the eigenvalues of $\frac{1}{n}\boldsymbol{X}^{\intercal}\boldsymbol{X}$ are bounded away from $0$ and $\infty$ with probability converging to one, and that there exists some constant $U$ that uniformly bounds the $L_2$-norm of the HiGLASSO estimator for all $n$.  If $a_n/\sqrt{n} \rightarrow \infty$, $a_n/n \rightarrow 0$, and $b_n \rightarrow 0$ as $n \rightarrow \infty$, then we have $P\big(\big\Vert\hat{\boldsymbol{\beta}}_{\mathcal{P}_1^{\mathsf{c}}}\big\Vert_2=0\big) \rightarrow 1$ and  $P\big(\big\Vert\hat{\boldsymbol{\gamma}}_{\mathcal{P}_2^{\mathsf{c}}}\big\Vert_2=0\big) \rightarrow 1$.\\

\noindent \textit{Proof:} See Web Appendix \textcolor{cyan}{B}.

The theorem ensures that spurious covariates will be eliminated by the HiGLASSO procedure when the number of covariates is fixed as $n \to \infty$. However, the theorem assumes conditions on the design matrix $\boldsymbol{X}$ which do not allow the number of covariates to diverge. Generalizations of sparsistency of the HiGLASSO estimator in high dimensional settings, i.e. when $|\mathcal{P} \cup \mathcal{P}^{\mathsf{c}}| = o(n)$, are not discussed here.

\section{Simulation study}
\label{simulation}
The goal of the simulation study is to compare the performance of HiGLASSO with alternative approaches for selecting main and pairwise interaction effects. The competing methods accounting for linear main effects and linear pairwise interaction terms include LASSO and hierNet. An alternative method accounting for nonlinear main effects and, potentially, nonlinear interaction terms is group LASSO. ``Nonlinear" in this context refers to nonlinear basis expansions of the original exposure variables. For the present simulation study we use a cubic basis expansion, where each scalar exposure variable $x_j$ is expanded to $(x_j, x_j^2, x_j^3)^{\top}$. Nonlinear interactions are therefore comprised of all pairwise multiples of individual terms in the corresponding basis expansions. For all methods with group structure, i.e. group LASSO and HiGLASSO, the full nonlinear basis expansions for each covariate define the groups ($p_s = 3$, $\forall s = 1, \cdots, S$). Similarly, for pairwise nonlinear interactions, all pairwise multiples of individual terms in the two basis expansions are considered a group. The \textsf{R} package \textsf{glmnet} was used to implement LASSO, the \textsf{R} package \textsf{hierNet} was used to implement hierNet, the \textsf{R} package \textsf{gglasso} was used to implement group LASSO, and the \textsf{R} package \textsf{higlasso} was used to implement HiGLASSO. VANISH was not considered in this simulation study because there is no publicly available implementation on CRAN.

\subsection{Simulation setting}
For the present simulation study, we consider 9 different scenarios, each with 500 simulated datasets and a sample size of either $n = 1000$ or $n = 10000$. The data generation mechanism for the simulated datasets is to first generate covariate vectors from a $N(0,\Sigma)$ distribution where $\Sigma$ is an compound symmetric matrix with unit variance and pairwise correlations equal to $0.3$ and then draw $\boldsymbol{y}|\boldsymbol{x}_1,...,\boldsymbol{x}_p$ from the regression model
\begin{equation*}
    \boldsymbol{y}=f(\boldsymbol{x}_1, \cdots, \boldsymbol{x}_p) + \boldsymbol{\epsilon}, \hspace{2 mm} \boldsymbol{\epsilon} \sim N(0,9\boldsymbol{I})
\end{equation*}
A list of the mean functions $(f(\cdot))$ and the number of predictors ($p = 10$, $p = 20$), across the six $n = 1000$ simulation scenarios are provided in Table \ref{simsetting}. The $n = 10000$ simulation settings have the same mean functions as the $n = 1000$ simulation settings, but were only considered with $p = 10$ in order to assess the large sample behavior of each method. In the `Scenario' column in Table \ref{simsetting}, L refers to scenarios with true linear main and interaction effects, PL refers to scenarios with true piecewise linear main and interaction effects, and NL refers to scenarios with true nonlinear main and interaction effects.

\begin{table}[!htbp]
\begin{center}
{
\resizebox{\columnwidth}{!}{
\begin{tabular}{ccc}
\hline
\textbf{Scenario} & $\boldsymbol{p}$ & \textbf{Mean Function}   \\ \hline
\rowcolor{lightgray}
 L10, L20 & 10, 20 & $\boldsymbol{x}_1 + \boldsymbol{x}_2 + \boldsymbol{x}_3 + \boldsymbol{x}_4 + \boldsymbol{x}_5 + \boldsymbol{x}_1\boldsymbol{x}_2 + \boldsymbol{x}_1\boldsymbol{x}_3 + \boldsymbol{x}_1\boldsymbol{x}_4 +$  \\
 \rowcolor{lightgray}
  & & $\boldsymbol{x}_1\boldsymbol{x}_5 + \boldsymbol{x}_2\boldsymbol{x}_3 + \boldsymbol{x}_2\boldsymbol{x}_4 + \boldsymbol{x}_2\boldsymbol{x}_5 + \boldsymbol{x}_3\boldsymbol{x}_4 + \boldsymbol{x}_3\boldsymbol{x}_5 + \boldsymbol{x}_4\boldsymbol{x}_5$ \\
  PL10, PL20 & 10, 20 & $\boldsymbol{x}_1I(\boldsymbol{x}_1 > 0) + \boldsymbol{x}_2I(\boldsymbol{x}_2 < 0) + \boldsymbol{x}_3I(\boldsymbol{x}_3 > 0.5) + \boldsymbol{x}_4I(\boldsymbol{x}_4 > 0) + \boldsymbol{x}_5I(\boldsymbol{x}_5 < -0.5) +$ \\
  & & $\boldsymbol{x}_1\boldsymbol{x}_2I(\boldsymbol{x}_1 > 0)I(\boldsymbol{x}_2 < 0) + \boldsymbol{x}_1\boldsymbol{x}_3I(\boldsymbol{x}_1 > 0)I(\boldsymbol{x}_3 > 0.5) + \boldsymbol{x}_1\boldsymbol{x}_4I(\boldsymbol{x}_1 > 0)I(\boldsymbol{x}_4 > 0) +$ \\
  & & $\boldsymbol{x}_1\boldsymbol{x}_5I(\boldsymbol{x}_1 > 0)I(\boldsymbol{x}_5 < -0.5) + \boldsymbol{x}_2\boldsymbol{x}_3I(\boldsymbol{x}_2 < 0)I(\boldsymbol{x}_3 > 0.5) + \boldsymbol{x}_2\boldsymbol{x}_4I(\boldsymbol{x}_2 < 0)I(\boldsymbol{x}_4 > 0)+$ \\
  & & $\boldsymbol{x}_2\boldsymbol{x}_5I(\boldsymbol{x}_2 < 0)I(\boldsymbol{x}_5 < -0.5) + \boldsymbol{x}_3\boldsymbol{x}_4I(\boldsymbol{x}_3 > 0.5)I(\boldsymbol{x}_4 > 0) +$ \\
  & & $\boldsymbol{x}_3\boldsymbol{x}_5I(\boldsymbol{x}_3 > 0.5)I(\boldsymbol{x}_5 < -0.5) + \boldsymbol{x}_4\boldsymbol{x}_5I(\boldsymbol{x}_4 > 0)I(\boldsymbol{x}_5 < -0.5)$ \\
\rowcolor{lightgray}
  NL10, NL20 & 10, 20 &  $\boldsymbol{x}_1I(\boldsymbol{x}_1 > 0) + \exp(\boldsymbol{x}_2) + |\boldsymbol{x}_3| + \boldsymbol{x}_4^2 + (\boldsymbol{x}_5+1)^2 + \boldsymbol{x}_1\exp(\boldsymbol{x}_2)I(\boldsymbol{x}_1 > 0)$  \\
\rowcolor{lightgray}
  & & $\boldsymbol{x}_1|\boldsymbol{x}_3|I(\boldsymbol{x}_1 > 0) + \boldsymbol{x}_1\boldsymbol{x}_4^2I(\boldsymbol{x}_1 > 0) + \boldsymbol{x}_1(\boldsymbol{x}_5+1)^2I(\boldsymbol{x}_1 > 0) + \exp(\boldsymbol{x}_2)|\boldsymbol{x}_3| +$ \\
\rowcolor{lightgray}
  & & $\exp(\boldsymbol{x}_2)\boldsymbol{x}_4^2 + \exp(\boldsymbol{x}_2)(\boldsymbol{x}_5+1)^2 + |\boldsymbol{x}_3|\boldsymbol{x}_4^2 + |\boldsymbol{x}_3|(\boldsymbol{x}_5+1)^2 + \boldsymbol{x}_4^2(\boldsymbol{x}_5+1)^2$ \\
  \hline
\end{tabular}
}}
\end{center}
\caption{Mean specifications for all simulation scenarios. In the scenario column, ``L'' indicates linear main and pairwise interaction effects, ``PL'' indicates piecewise linear main and interaction effects, and ``NL'' indicates nonlinear main and interaction effects. $p$ represents the number of predictors.}
\label{simsetting}
\end{table}

If we consider the cubic spline expansion with all possible two-way interactions, $p=10$ corresponds to 435 effective predictors in our design matrix and $p=20$ corresponds to 1770 effective predictors in our design matrix. Tuning parameters for each regularized regression method are selected via 10-fold cross-validation. For LASSO, group LASSO, and hierNet, the largest tuning parameter value within one standard error of the minimum cross-validation error is selected. Since HiGLASSO is naturally conservative with respect to interaction selection, the tuning parameter pair that results in the lowest cross-validation error is selected. With these tuning parameter values, the corresponding regularized regression methods are then re-fit on the full data.

\subsection{Performance metrics}
\label{performance}
The simulation metrics that we will focus on are the following:
\begin{enumerate}
    \item \textbf{\underline{False negative main effects rate (FNM)}}: The average number of times that a non-null main effect term is not selected by a model.
    \item \textbf{\underline{False positive main effects rate (FPM)}}: The average number of times that a null main effect term is selected by a model.
    \item \textbf{\underline{False negative interaction effects rate (FNI)}}: The average number of times that a non-null interaction effect term is not selected by a model.
    \item \textbf{\underline{False positive interaction effects rate (FPI)}}: The average number of times that a null interaction effect term is selected by a model.
\end{enumerate}
These four metrics are scaled to a range between 0 and 100, reflecting the average percent error rate per simulated data set and per important/unimportant term. Note that smaller values of all four metrics indicate better variable selection performance.

\subsection{Simulation results}

\begin{figure}[!htbp]
\centering
\scalebox{0.8}{\includegraphics[width=1.0\columnwidth]{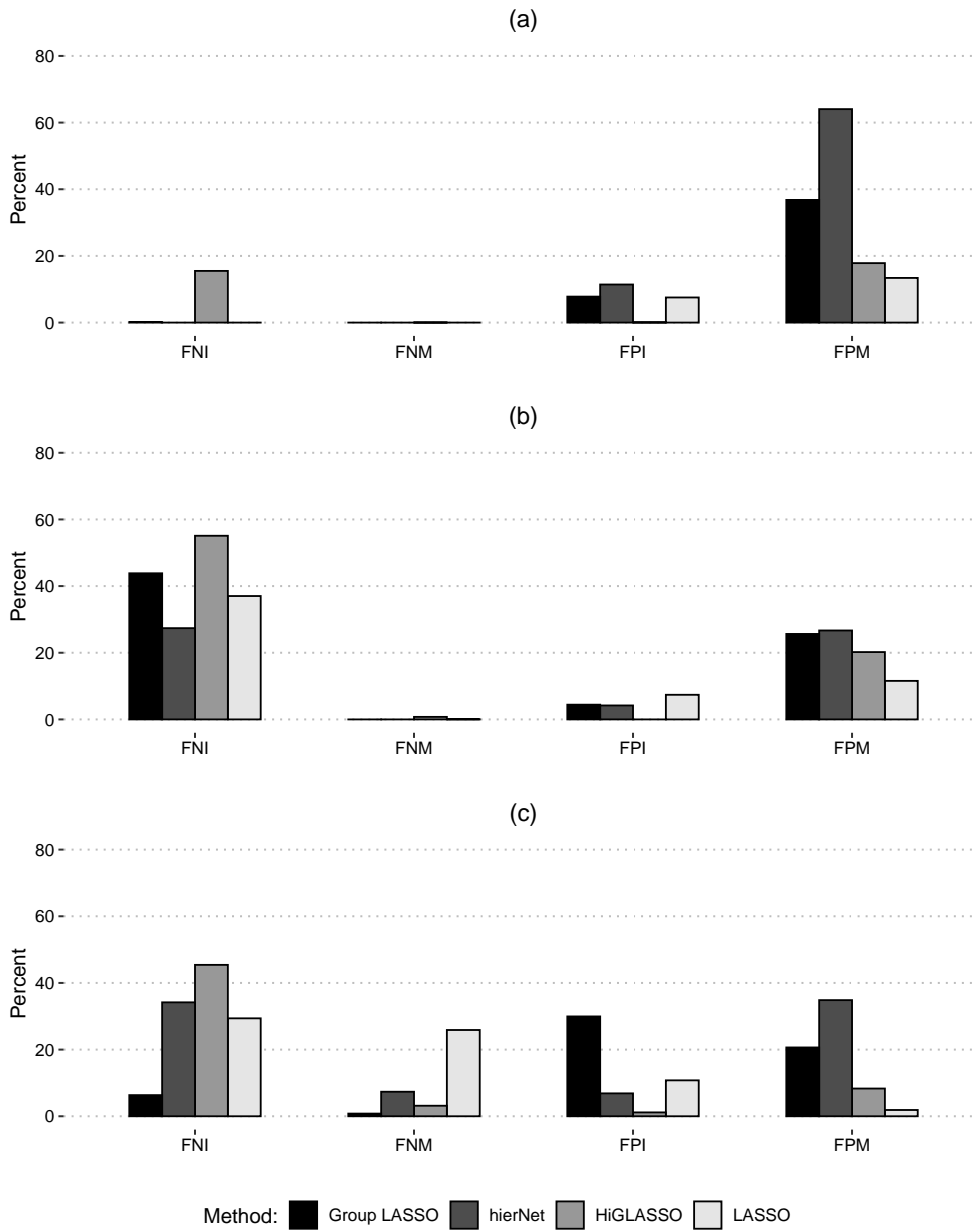}}
\caption{Simulation Results for the $n = 1000$ and $p = 10$ cases: (a) L10 (b) PL10 (c) NL10. FNI, FNM, FPI, and FPM are defined in Section \ref{performance}.}
\label{HiGLASSO_simulation_n1000_p10}
\end{figure}

\begin{figure}[!htbp]
\centering
\scalebox{0.8}{\includegraphics[width=1.0\columnwidth]{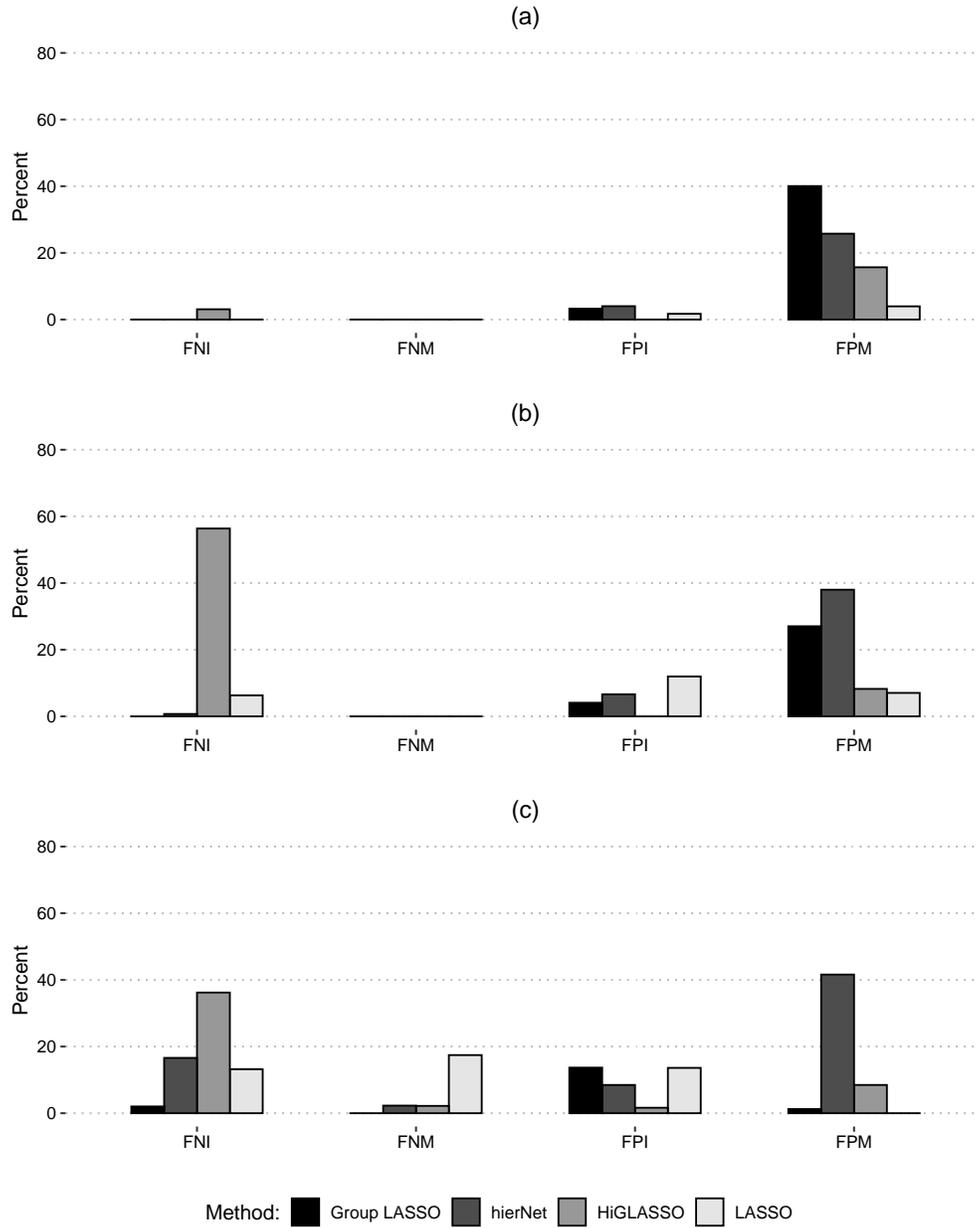}}
\caption{Simulation Results for the $n = 10000$ and $p = 10$ cases: (a) Linear main and interaction effects (b) Piecewise linear main and interaction effects (c) Nonlinear main and interaction effects. FNI, FNM, FPI, and FPM are defined in Section \textcolor{red}{4.2}.}
\label{HiGLASSO_simulation_n10000}
\end{figure}

Simulation results for the $n = 1000$ and $p = 10$ simulation scenarios are presented in Figure \ref{HiGLASSO_simulation_n1000_p10}. Panel (a) corresponds to case L10 with linear main and interaction effects, panel (b) corresponds to case PL10 with piecewise linear main and interaction effects, and panel (c) corresponds to case NL10 with nonlinear main and interaction effects (see Figure \ref{HiGLASSO_simulation_n10000} for the $n = 10000$ simulation results). In L10, LASSO is correctly specified, and therefore leads to relatively low FNI, FNM, FPI, and FPM. LASSO's FNM, FPI, and FPM in PL10 are comparable to the respective metrics in L10, however the FNI is notably larger (FNI = 37\%). For NL10, some of the main effects contain absolute values and quadratic terms, which are more difficult for LASSO with only linear main and interaction terms to detect, hence the elevated FNI (FNI = 29\%) and FNM (FNM = 26\%). hierNet tends to do well with respect to FNI, FNM, and FPI, but on average has the highest FPM for L10 (FPM = 64\%), PL10 (FPM = 27\%), and NL10 (FPM = 35\%). Conversely, HiGLASSO has the highest FNI rate for L10 (FNI = 16\%), PL10 (FNI 55\%), and NL10 (FNI = 45\%), but has relatively low FNM, FPI, and FPM. That is, HiGLASSO is conservative for interaction selection, but when HiGLASSO selects interactions, they are almost always true interactions. Group LASSO's behavior is difficult to characterize across the three simulation scenarios. One general theme is that the FPM for group LASSO is above 20\% for L10, PL10, and NL10. Group LASSO also has an FNI of 44\% for P10. The FNI, FNM, FPI, and FPM patterns for the $n = 10000$ and $p = 10$ simulation scenarios are similar to the $n = 1000$ and $p = 10$ simulation scenarios, however there is a general decrease in false negative and false positive rates across all methods.

\begin{figure}[!htbp]
\centering
\scalebox{0.8}{\includegraphics[width=1.0\columnwidth]{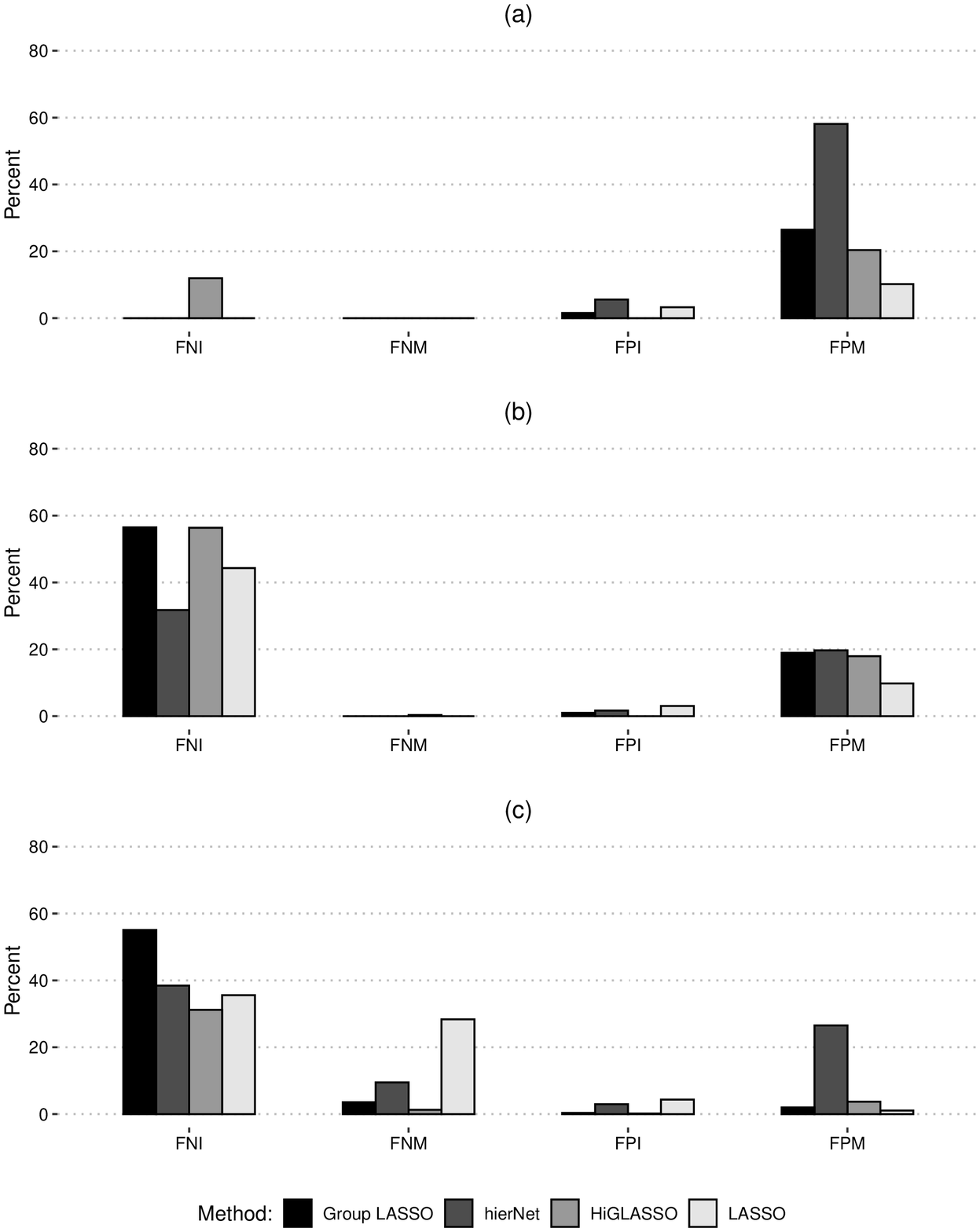}}
\caption{Simulation Results for the $n = 1000$ and $p = 20$ cases: (a) L20 (b) PL20 (c) NL20. FNI, FNM, FPI, and FPM are defined in Section \ref{performance}.}
\label{HiGLASSO_simulation_p20}
\end{figure}

Figure \ref{HiGLASSO_simulation_p20} summarizes the simulation results for $n = 1000$ and $p = 20$ simulation scenarios. Panel (a) corresponds to case L20 with linear main and interaction effects, panel (b) corresponds to case PL20 with piecewise linear main and interaction effects, and panel (c) corresponds to case NL20 with nonlinear main and interaction effects. Simulation results for L20 and PL20 are nearly identical to the simulation results for L10 and L20, however the simulation results for NL20 are different from the simulation results for NL10. The notable difference in NL20 is that HiGLASSO now has the lowest FNI (FNI = 31\%), FNM (FNM = 1\%), and FPI (FPI = 0.2\%), but also has very low FPM (FPM = 4\%). For NL20, hierNet maintains an elevated FPM (FPM = 27\%), LASSO has increased false negative rates (FNI = 36\%, FNM = 28\%), and group LASSO has a large FNI (FNI = 55\%) as opposed to the higher false positive rates from NL10.

When there are nonlinear main and interaction effects in the true exposure-response model such that the nonlinear interactions obey the strong heredity principle, HiGLASSO has excellent performance with respect to FNM, FPI, and FPM. HiGLASSO can be conservative for interaction selection, as evidenced by elevated FNI in Figure \ref{HiGLASSO_simulation_n1000_p10} and Figure \ref{HiGLASSO_simulation_p20}, for which there are several explanations. When the true outcome-exposure association involves sufficiently linear main and interaction effects, HiGLASSO overparameterizes the exposure-response model and therefore unnecessarily introduces additional parameters that require estimation. Estimating the additional parameters results in a loss of power to detect all of the true interactions (although the false discovery rate for main and interaction effects is very low). Another explanation is that using a cubic basis expansion to handle nonlinear main and interaction effects involves a certain level of approximation error. Nevertheless, HiGLASSO shows great promise in the NL20 setting, which is the scenario that it is specifically designed for.

\section{Application to the LIFECODES study}
\label{BWH}
\subsection{Data overview}
LIFECODES is a biobank that longitudinally collects biospecimens and medical data across pregnancy with the two-part goal of (a) understanding biophysiological processes underlying fetal development and (b) identifying environmental risk factors for adverse birth outcomes. A subset of pregnant women in the LIFECODES cohort ($n=482$) had 21 phthalate, phenol, and paraben concentrations (see Table \ref{BWHexposure}) measured longitudinally at approximately, 10 weeks, 18 weeks, 26 weeks, and 35 weeks gestation. Due to known temporal variability in the analytes of interest, specific gravity adjusted geometric averages across the first three visits for each contaminant and each subject were used as covariates to minimize measurement error \cite{meeker2012}. The fourth visit measurement was omitted because many women with preterm deliveries had already delivered by 35 weeks. Of those $482$ women, our working dataset contains $n=477$ women ($128$ preterm deliveries and $349$ full-term deliveries) after removal of subjects with no phenol measurements. Study details including exclusion criteria, handling and storage of biological samples, assessment of contaminant concentrations, and institutional review board approval can all be found in \cite{ferguson2015b}.

In this section, we apply LASSO, group LASSO, hierNet, and HiGLASSO to the data collected as a part of LIFECODES where the covariates are the 21 phthalate, paraben, and phenol geometric averages (log-transformed and standardized) and the outcome is specific gravity corrected 8-isoprostane, a biomarker that is indicative of oxidative stress, averaged over the first three visits (log-transformed and centered) \cite{montuschi1999}. For the nonlinear methods we expand each of the 21 exposure variables into a group of two variables using a quadratic basis expansion.

\begin{table}[!htbp]
    \begin{center}{
    \resizebox{0.6\columnwidth}{!}{
    \begin{tabular}{lll}
         \hline
         \textbf{Exposure Class} & \textbf{Full Name} & \textbf{Acronym} \\
         \hline
         Phthalates & & \\
         & mono-n-butyl & MBP \\
         & monobenzyl & MBzP \\
         & mono(3-carboxypropyl) & MCPP \\
         & mono(2-ethyl-5-carboxypentyl) & MECPP \\
         & mono(2-ethyl-5-hydroxyhexyl) & MEHHP \\
         & mono(2-ethylhexyl) & MEHP \\
         & mono(2-ethyl-5-oxohexyl) & MEOHP \\
         & monoethyl & MEP \\
         & monoisobutyl & MiBP \\
         & Summed di(2-ethylhexyl) & DEHP \\
         Phenols & & \\
         & 2,4-Dichlorophenol & 2,4-DCP \\
         & 2,5-Dichlorophenol & 2,5-DCP \\
         & benzophenone-3 & BP3 \\
         & Bisphenol A & BPA \\
         & Bisphenol S & BPS \\
         & butyl paraben & BuPB \\
         & ethyl paraben & EtPB \\
         & methyl paraben & MePB \\
         & propyl paraben & PrPB \\
         & triclocarban & TCC \\
         & triclosan & TCS \\
         \hline
    \end{tabular}}}
    \end{center}
    \caption{List of 21 exposure measurements including 10 phthalates and 11 phenols in the LIFECODES dataset.}
    \label{BWHexposure}
\end{table}

\subsection{Initial analyses}
\begin{figure}[!htbp]
\centering
\scalebox{0.8}{\includegraphics[width=1.0\columnwidth]{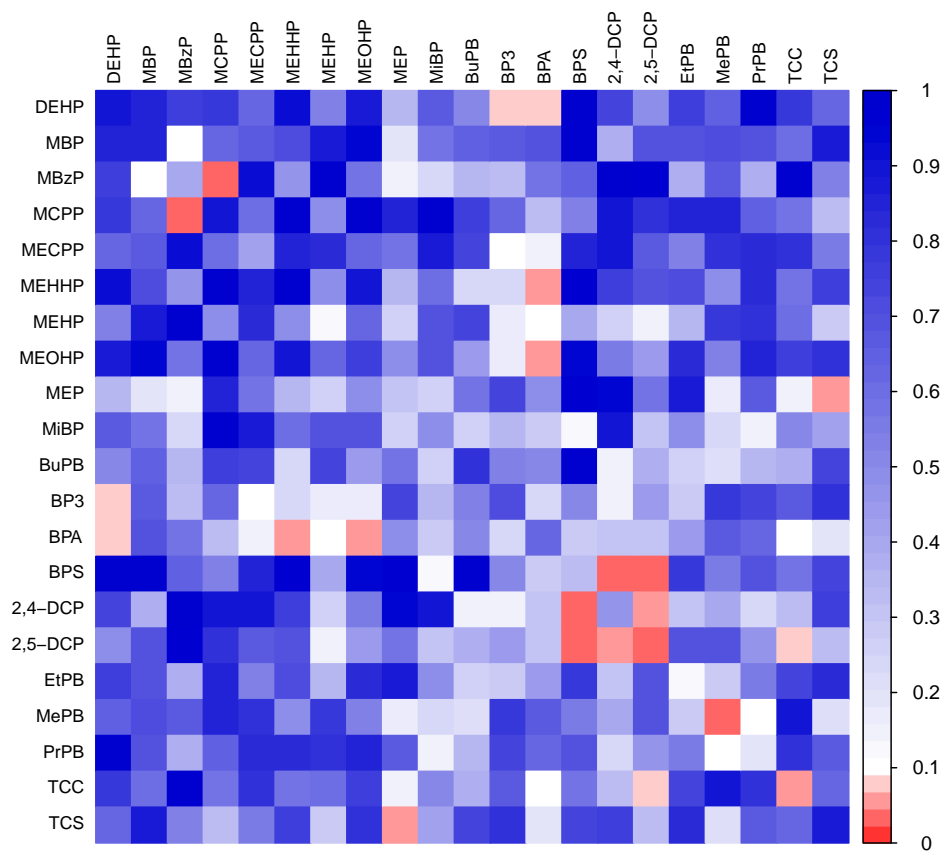}}
\caption{Heatmap for Wald test p-values corresponding to all pairwise linear interactions. Each p-value is obtained from a multiple regression model with 21 exposure main effect terms and a single pairwise linear interaction term. Diagonal elements indicate the addition of a squared term instead of an interaction.}
\label{BWH_heatmap_supp}
\end{figure}

To perform an interaction search, many analysts will proceed by adding linear pairwise interaction terms one-at-a-time and then subsequently assess the statistical significance of each interaction. Therefore, as a cursory analytical step, we will regress log-transformed 8-isoprostane on every possible linear pairwise interaction term one at a time, keeping the 21 linear main effects for each exposure in the model throughout. Figure \ref{BWH_heatmap_supp} provides a visualization of the resulting p-values for each pairwise interaction (diagonal entries of the heatmap are p-values for the addition of a squared term in the linear regression model). We observe that there are several interactions that fall below the $p < 0.05$ threshold, including MBzP$\times$MCPP ($p = 0.026$), BPS$\times$2,4-DCP ($p = 0.025$), and BPS$\times$2,5-DCP ($p = 0.016$). Moreover, the Wald tests for inclusion of a 2,5-DCP squared term ($p = 0.015$) and MePB squared term ($p = 0.033$) are significant at the $\alpha = 0.05$ level. Lastly, looking at the unadjusted, marginal exposure-response associations we can clearly identify several nonlinear relationships (see Figure \ref{BWH_nonlinear}). These exploratory steps affirm that a model accounting for nonlinearity and interaction structure in the exposure-response surface may be desired.

\subsection{Variable selection results}

\begin{figure}[!htbp]
\centering
\scalebox{0.8}{\includegraphics[width=1.0\columnwidth]{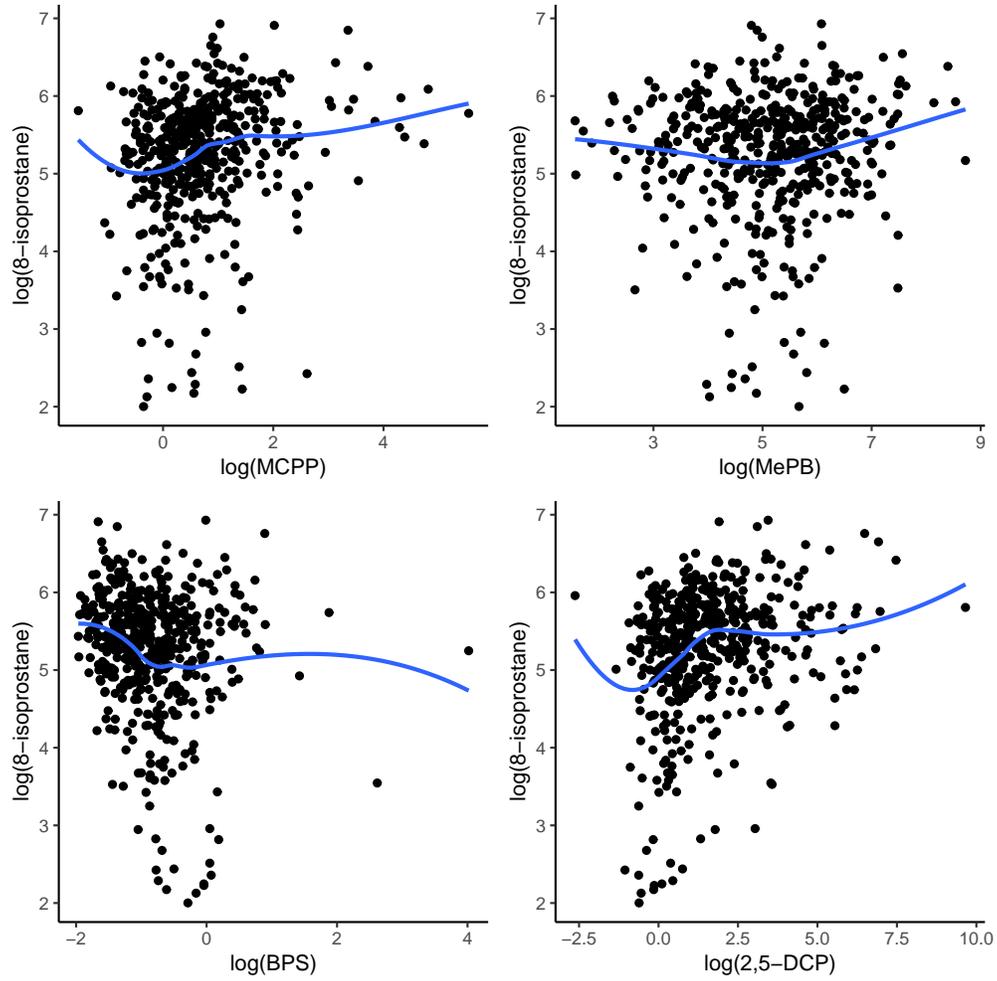}}
\caption{Scatterplots between four exposures and 8-isoprostane superimposed with a Locally Weighted Scatterplot Smoothing (LOWESS) curve. The four exposures are mono(3-carboxypropyl) phthalate (MCPP), methyl paraben (MePB), Bisphenol S (BPS), and 2,5-Dichlorophenol (2,5-DCP).}
\label{BWH_nonlinear}
\end{figure}

The selected main effects and interaction effects for each method are enumerated in Table \ref{BWHselection}. The two methods that only account for linear pairwise interaction effects, LASSO and hierNet, have very similar results. Namely, all 9 main effects and 2 out of 3 interaction effects selected by LASSO are also selected by hierNet. The one interaction that is selected by LASSO but not hierNet is MEP$\times$TCS, which violates strong heredity. There are more main effects selected by group LASSO than any other method. Moreover, the set of main effects selected by all other methods is a proper subset of the main effects selected by group LASSO. However, 4 of the 6 interactions selected by group LASSO violate strong heredity, the only exceptions being MBzP$\times$MCPP and BPS$\times$2,5-DCP. HiGLASSO selects fewer main effects and interaction effects than group LASSO, but the interactions both satisfy strong heredity. In fact, the interactions selected by HiGLASSO are MBzP$\times$MCPP and BPS$\times$2,5-DCP, which is consistent with group LASSO. One other interesting observation is that group LASSO and HiGLASSO both select MePB, while LASSO and hierNet do not. Referring to Figure \ref{BWH_nonlinear}, we can visually identify a marginal quadratic relationship between MePB (log-transformed) and 8-isoprostane (log-transformed), which when modeled by a linear term would be relatively flat. The quadratic term in the basis expansion facilitates detection of an association between MePB and 8-isoprostane that would have been missed otherwise.

\begin{table}[!htbp]
\begin{center}
{
\resizebox{0.7\columnwidth}{!}{
\rowcolors{1}{}{lightgray}
\begin{tabular}{ccccc}
\hline
\textbf{Selected Term} & \textbf{LASSO} & \textbf{hierNet} & \textbf{Group LASSO} & \textbf{HiGLASSO} \\ \hline
MBP & \checkmark & \checkmark & \checkmark & \\
MBzP & \checkmark & \checkmark & \checkmark & \checkmark \\
MCPP & & \checkmark & \checkmark & \checkmark \\
MECPP & \checkmark & \checkmark & \checkmark & \checkmark \\
MEP & \checkmark & \checkmark & \checkmark & \checkmark \\
MiBP & \checkmark & \checkmark & \checkmark & \\
BuBP & \checkmark & \checkmark & \checkmark & \checkmark \\
BPS & \checkmark & \checkmark & \checkmark & \checkmark \\
2,5-DCP & \checkmark & \checkmark & \checkmark & \checkmark \\
EtPB & & & \checkmark & \\
MePB & & & \checkmark & \checkmark \\
TCC & \checkmark & \checkmark & \checkmark & \checkmark \\
MBP$\times$BPA & & & \checkmark & \\
MBP$\times$MBzP & & \checkmark & & \\
MBP$\times$MCPP & \checkmark & \checkmark & & \\
MBzP$\times$MCPP & \checkmark & \checkmark & \checkmark & \checkmark \\
MECPP$\times$BP3 & & & \checkmark & \\
MECPP$\times$BPA & & & \checkmark & \\
MEP$\times$TCS & \checkmark & & & \\
MiBP$\times$MBzP & & \checkmark & & \\
BP3$\times$BPA & & & \checkmark & \\
BPS$\times$2,5-DCP & & & \checkmark & \checkmark \\
\hline
\end{tabular}
}}
\end{center}
\caption{Selected main effects and interaction effects from the LIFECODES study. Candidate main and interaction effects that were not selected are omitted for brevity.}
\label{BWHselection}
\end{table}

\section{Discussion}
\label{discussion}
This paper presents a new penalized variable selection algorithm to handle groups or sets of correlated predictors and their possibly nonlinear interactions. HiGLASSO imposes strong heredity, induces sparsity as in group LASSO, and maintains efficiency and sparsistency through the use of integrative weights. The integrative weights in HiGLASSO also help select a more parsimonious model compared to other penalized regression strategies, as seen in the LIFECODES data example. By defining groups through basis expansions, the method can handle nonlinear main effects and nonlinear pairwise interactions. Our simulation results indicate that HiGLASSO controls false discovery rates while having competitive true discovery rates for both main effects and interactions, particularly when there is true nonlinearity in the exposure-response surface. Further extension of HiGLASSO to an elastic-net framework is needed in order to handle highly collinear groups of environmental exposures. Principled post model selection inference and robust replication strategies are other areas following such initial interaction screening strategies that require rapid development.

Because exposures never occur in isolation, identifying exposure interactions is crucial to advancing the understanding of how the environment holistically influences health. We show that non-linearity in exposure-response associations and interactions, a common feature in epidemiologic studies, can make these effects difficult to quantify. HiGLASSO is useful in this space as a pairwise interaction detection tool that can help identify possibly nonlinear interaction effects and, ultimately, advance research on environmental chemical mixtures beyond models that strictly assume additive exposure effects or linear interaction effects.

\section*{Acknowledgements}

Research reported in this publication was supported by NIH grant ES 20811 (BM), NSF grant 1712933 (BM), NSF DMS 1811768 (NN), NIH grant R01ES018872 (BM and JM), NIH Grant P42ES017198 (BM and JM), and NIH grant UH3OD023251 (BM and JM). Funding for ZW is provided by the National Cancer Institute of the National Institutes of Health under award number P30CA046592 (Cancer Center Support Grant (CCSG) Development Funds from Rogel Cancer Center). Funding for KF was provided by the Intramural Research Program of the National Institute of Environmental Health Sciences, National Institutes of Health.\vspace*{-8pt}

\bibliographystyle{unsrt}  
\bibliography{main}  

%
%
%
%

\newpage

\section*{Web Appendix A: HiGLASSO algorithm}
\label{alg}

\subsection*{\normalfont{A.1.} \textit{Objective Function}}

The HiGLASSO objective function is:

\begin{align}
 & \underset{\boldsymbol{\beta}_j, \boldsymbol{\eta}_{jj'}}{\arg\min} \hspace{1 mm} \frac{1}{2}\Big\Vert\boldsymbol{y}-\sum_{j=1}^{S}\boldsymbol{X}_j\boldsymbol{\beta}_j-\sum_{1 \leq j<j' \leq S}\boldsymbol{X}_{jj'}[\boldsymbol{\eta}_{jj'} \odot (\boldsymbol{\beta}_j \otimes \boldsymbol{\beta}_{j'})]\Big\Vert_2^2 & \label{HiGLASSOcriterion} \\
  & + \lambda_1\sum_{j = 1}^{S}w_j||\boldsymbol{\beta}_j||_2 + \lambda_2\sum_{1 \leq j<j' \leq S}w_{jj'}||\boldsymbol{\eta}_{jj'}||_2, \nonumber &
\end{align}

\begin{equation}
\label{weightfn1}
    w_j \equiv \text{exp}\bigg\{-\frac{||\boldsymbol{\beta}_j||_{\infty}}{\sigma}\bigg\} \text{ for } j = 1, \cdots, S,
\end{equation}
\begin{equation}
\label{weightfn2}
w_{jj'} \equiv \text{exp}\bigg\{-\frac{||\boldsymbol{\eta}_{jj'}||_{\infty}}{\sigma}\bigg\} \text{ for } 1 \leq j < j' \leq S,
\end{equation}

\subsection*{\normalfont{A.2.} \textit{Updating main effect coefficients}}
\label{updatemain}
By substituting our weight function (\ref{weightfn1}) into (\ref{HiGLASSOcriterion}), given the current $\hat{\boldsymbol{\beta}}_{j'}$'s with $j' \neq j$ and $\hat{\boldsymbol{\eta}}_{jj'}$'s, the objective function can be written as
\begin{equation}
\label{HiGLASSOcriterionmain}
\underset{\boldsymbol{\beta}_j}{\arg\min} \hspace{1 mm} \frac{1}{2}\big\Vert\tilde{\boldsymbol{y}}-\tilde{\boldsymbol{X}}_j\boldsymbol{\beta}_j\big\Vert^2_2 + \lambda_1\text{exp}\bigg\{-\frac{||\boldsymbol{\beta}_j||_{\infty}}{\sigma}\bigg\}||\boldsymbol{\beta}_j||_2,
\end{equation}
such that
\begin{equation*}
\tilde{\boldsymbol{y}}=\boldsymbol{y}-\sum_{k \neq j}\boldsymbol{X}_k\hat{\boldsymbol{\beta}}_k-\sum_{k, l \neq j}\boldsymbol{X}_{kl}[\hat{\boldsymbol{\eta}}_{kl} \odot (\hat{\boldsymbol{\beta}}_k \otimes \hat{\boldsymbol{\beta}}_{l})],
\end{equation*}
\begin{equation*}
\tilde{\boldsymbol{X}}_j=\boldsymbol{X}_j+\sum_{k<j}\boldsymbol{X}_{kj}\cdot\text{diag}(\hat{\boldsymbol{\eta}}_{kj})(\hat{\boldsymbol{\beta}}_k \otimes \boldsymbol{I}_{p_j})+\sum_{l>j}\boldsymbol{X}_{jl}\cdot\text{diag}(\hat{\boldsymbol{\eta}}_{jl})(\boldsymbol{I}_{p_j} \otimes \hat{\boldsymbol{\beta}}_l),
\end{equation*}
where $\boldsymbol{I}_{p_j}$ is $p_j$ dimensional identity matrix. $\tilde{\boldsymbol{X}}_j$ and $\tilde{\boldsymbol{y}}$ represent the design matrix and response vector at current step. (\ref{HiGLASSOcriterionmain}) can be directly solved using gradient descent or the Newton-Raphson algorithm \cite{bauer2009}.

Alternatively, we obtain updating algorithm for $\boldsymbol{\beta}_j$ in closed form using local quadratic approximation (LQA) \cite{fan2001}. Let \textbf{Pen}$_1(\boldsymbol{\beta}_j)$ denote the penalty term in (\ref{HiGLASSOcriterionmain}). We approximate \textbf{Pen}$_1(\boldsymbol{\beta}_j)$ by
\begin{equation*}
    \textbf{Pen}_1(\boldsymbol{\beta}_j) \approx \textbf{Pen}_1\Big(\hat{\boldsymbol{\beta}}_j^{(m)}\Big) + \frac{1}{2}\sum_{k=1}^{p_j}d_{jk}^{(m)}\bigg[\beta_{jk}^2-\Big(\hat{\beta}_{jk}^{(m)}\Big)^2\bigg]
\end{equation*}
where $\beta_{jk}$ is the $k^{th}$ element of $\boldsymbol{\beta}_j$, $\hat{\boldsymbol{\beta}}_j^{(m)}$ is the estimate of $\boldsymbol{\beta}_j$ from $m^{th}$ iteration, and $d_{jk}$ is defined through $$\frac{\partial \textbf{Pen}_1(\boldsymbol{\beta}_j)}{\partial \beta_{jk}}=d_{jk}\beta_{jk}.$$
By calculating the derivative of \textbf{Pen}$_1(\boldsymbol{\beta}_j)$, we have
\begin{align}
 d_{jk} &=
  \begin{cases}
   \text{exp}\Big\{-\frac{||\boldsymbol{\beta}_j||_{\infty}}{\sigma}\Big\}\big(||\boldsymbol{\beta}_j||_2\big)^{-1},        & \text{if } |\beta_{jk}| \neq ||\boldsymbol{\beta}_j||_{\infty}  \\
   \text{exp}\Big\{-\frac{||\boldsymbol{\beta}_j||_{\infty}}{\sigma}\Big\}\Big[\big(||\boldsymbol{\beta}_j||_2\big)^{-1}-||\boldsymbol{\beta}_j||_2\big(|\beta_{jk}|\sigma\big)^{-1}\Big], & \text{if } |\beta_{jk}| = ||\boldsymbol{\beta}_j||_{\infty}.
  \end{cases}
  \label{pen1d}
\end{align}
The problem with LQA is that $d_{jk}$, which represents the second-degree derivative of $\textbf{Pen}_1(\boldsymbol{\beta}_j)$, might be negative when $|\beta_{jk}| = ||\boldsymbol{\beta}_j||_{\infty}$. Therefore, it is not guaranteed that the approximated \textbf{Pen}$_j(\boldsymbol{\beta}_j)$ will be convex.

Pan and Zhao proposed generalized local quadratic approximation (GLQA) to employ convex quadratic approximation to the penalty function \cite{pan2016}. Let $\mathcal{P}_1(\boldsymbol{\beta}_j)$ denote GLQA of \textbf{Pen}$_1(\boldsymbol{\beta}_j)$ that satisfies the following three properties
\begin{enumerate}
\item $\mathcal{P}_1(\boldsymbol{\beta}_j)$ is convex,
\vspace{2 mm}
\item $\mathcal{P}_1\Big(\hat{\boldsymbol{\beta}}_j^{(m)}\Big)=\textbf{Pen}_1\Big(\hat{\boldsymbol{\beta}}_j^{(m)}\Big)$,
\vspace{2 mm}
\item $\frac{\partial \mathcal{P}_1(\boldsymbol{\beta}_j)}{\partial \beta_{jk}}\Big|_{\beta_{jk}=\hat{\beta}_{jk}^{(m)}}=\frac{\partial \textbf{Pen}_1(\boldsymbol{\beta}_j)}{\partial \beta_{jk}}\Big|_{\beta_{jk}=\hat{\beta}_{jk}^{(m)}}$ $\forall$ $k$.
\end{enumerate}
A simple choice takes the form of
\begin{equation*}
    \mathcal{P}_1(\boldsymbol{\beta}_j)=\textbf{Pen}_1\Big(\hat{\boldsymbol{\beta}}_j^{(m)}\Big) + \frac{1}{2}\sum_{k = 1}^{p_j}\big|d_{jk}^{(m)}\big|\big[(\beta_{jk}^2 + c_1)^2 + c_2\big].
\end{equation*}
Solving $c_1$ and $c_2$ according to the second and third conditions gives
\begin{equation*}
    \mathcal{P}_1(\boldsymbol{\beta}_j)=\textbf{Pen}_1\Big(\hat{\boldsymbol{\beta}}_j^{(m)}\Big) +
    \frac{1}{2}\sum_{k = 1}^{p_j}\big|d_{jk}^{(m)}\big|\Bigg[\Bigg(\beta_{jk}^2 - \Bigg(1 - \frac{d_{jk}^{(m)}}{|d_{jk}^{(m)}|}\Bigg)\hat{\beta}_{jk}^{(m)}\Bigg)^2 - \Big(\hat{\beta}_{jk}^{(m)}\Big)^2\Bigg].
\end{equation*}
Rewriting the $\mathcal{P}_1(\boldsymbol{\beta}_j)$ in matrix form,  (\ref{HiGLASSOcriterionmain}) can be approximated as
\begin{equation*}
    \frac{1}{2}||\tilde{\boldsymbol{y}}-\tilde{\boldsymbol{X}}_j\boldsymbol{\beta}_j||^2_2 + \frac{1}{2}\lambda_1 \boldsymbol{\beta}_j^{\top} \boldsymbol{D}_j^{(m)} \boldsymbol{\beta}_j - \lambda_1 \boldsymbol{c}^{(m)}{}^{\top} \boldsymbol{\beta}_j + \text{Constant}
\end{equation*}
where $$\boldsymbol{D}_j^{(m)}=\text{diag}\Big[\Big(d_{j1}^{(m)}, \cdots, d_{jp_j}^{(m)}\Big)\Big] \text{ and }$$

$$\boldsymbol{c}^{(m)}=\Big\{\Big(\big|d_{j1}^{(m)}\big|-d_{j1}^{(m)}\Big)\hat{\beta}_{j1}^{(m)}, \cdots, \Big(\big|d_{jp_j}^{(m)}\big|-d_{jp_j}^{(m)}\Big)\hat{\beta}_{jp_j}^{(m)}\Big\}^{\top}.$$ $\boldsymbol{\beta}_j$ can be updated in closed-form as
\begin{equation}
\label{closeupdatebeta}
    \hat{\boldsymbol{\beta}}_j=\Big(\tilde{\boldsymbol{X}}_j^{\top}\tilde{\boldsymbol{X}}_j+n\lambda_1\boldsymbol{D}_j^{(m)}\Big)^{-1}\Big(\tilde{\boldsymbol{X}}_j^{\top}\tilde{\boldsymbol{y}}+\lambda_1 \cdot \boldsymbol{c}^{(m)}\Big).
\end{equation}

\subsection*{\normalfont{A.3.} \textit{Updating scalar terms associated with interactions}}
\label{updateinteraction}
By substituting the specified weight function (\ref{weightfn2}) into (\ref{HiGLASSOcriterion}), given $\hat{\boldsymbol{\beta}}_{j}$'s, the objective function can be expressed as
\begin{equation}
\label{HiGLASSOcriterionint}
\underset{\boldsymbol{\eta}_{jj'}}{\arg\min} \hspace{1 mm} \frac{1}{2}\Big\Vert\tilde{\boldsymbol{y}}-\sum_{j<j'}\tilde{\boldsymbol{X}}_{jj'}\boldsymbol{\eta}_{jj'}\Big\Vert_2^2 + \lambda_2\sum_{j<j'}\text{exp}\bigg\{-\frac{||\boldsymbol{\eta}_{jj'}||_{\infty}}{\sigma}\bigg\}||\boldsymbol{\eta}_{jj'}||_2
\end{equation}
where
\begin{equation*}
\tilde{\boldsymbol{y}}=\boldsymbol{y}-\sum_{k = 1}^{S} \boldsymbol{X}_k\hat{\boldsymbol{\beta}}_k
\end{equation*}
and
\begin{equation*}
\tilde{\boldsymbol{X}}_{jj'}=\boldsymbol{X}_{jj'}\text{diag}\big[(\hat{\boldsymbol{\beta}}_{j} \otimes \hat{\boldsymbol{\beta}}_{j'})\big] \text{ for } 1 \leq j < j' \leq S.
\end{equation*}

Let \textbf{Pen}$_2(\boldsymbol{\eta}_{jj'})$ denote the individual penalty term in (\ref{HiGLASSOcriterionint}) and let $\mathcal{P}_2(\boldsymbol{\beta}_{jj'})$ denote GLQA of \textbf{Pen}$_2(\boldsymbol{\eta}_{jj'})$. We have
\begin{equation*}
    \mathcal{P}_2(\boldsymbol{\eta}_{jj'})=\textbf{Pen}_1\Big(\hat{\boldsymbol{\eta}}_{jj'}^{(m)}\Big) +
    \frac{1}{2}\sum_{k = 1}^{p_jp_{j'}}\big|d_{jj'k}^{(m)}\big|\Bigg[\Bigg(\eta_{jj'k}^2 - \Bigg(1 - \frac{d_{jj'k}^{(m)}}{\big|d_{jj'k}^{(m)}\big|}\Bigg)\hat{\eta}_{jj'k}^{(m)}\Bigg)^2 - \Big(\hat{\eta}_{jj'k}^{(m)}\Big)^2\Bigg]
\end{equation*}
where $\eta_{jj'k}$ is the $k^{th}$ element of $(p_jp_{j'})-$vector of $\boldsymbol{\eta}_{jj'}$ and $d_{jj'k}$ is similarly defined through $$\frac{\partial \textbf{Pen}_2(\boldsymbol{\eta}_{jj'})}{\partial \eta_{jj'k}}=d_{jj'k}\eta_{jj'k}$$ as (\ref{pen1d}). (\ref{HiGLASSOcriterionint}) can be approximated as
\begin{equation*}
    \frac{1}{2}||\tilde{\boldsymbol{y}}-\tilde{\boldsymbol{X}}\boldsymbol{\eta}||^2_2 + \frac{1}{2}\lambda_2 \boldsymbol{\eta}^{\top} \boldsymbol{D}^{(m)} \boldsymbol{\eta} - \lambda_2 \boldsymbol{C}^{(m)}{}^{\top} \boldsymbol{\eta} + \text{Constant}
\end{equation*}
where $\tilde{\boldsymbol{X}}=[\tilde{\boldsymbol{X}}_{12}, \cdots, \tilde{\boldsymbol{X}}_{S-1, S}]$, $\boldsymbol{\eta}=\big(\boldsymbol{\eta}_{12}^{\top}, \cdots, \boldsymbol{\eta}_{S-1, S}^{\top}\big)^{\top}$, \\
$$\boldsymbol{D}^{(m)}=\text{diag}\Big[d_{121}^{(m)}, \cdots, d_{12(p_1p_2)}^{(m)}, \cdots, d_{(S-1)S(p_{S-1}p_S)}^{(m)}\Big],$$ and $\boldsymbol{C}^{(m)}$ is a $[S(S-1)/2] \times [\sum_{j<j'}p_jp_{j'}]$ block column vector such that the block corresponding to the interaction between group $j$ and group $j'$ is defined as a vector of length $p_jp_{j'}$ with the $k^{th}$ element equal to $\Big(\big|d_{jj'k}^{(m)}\big|-d_{jj'k}^{(m)}\Big)\hat{\eta}_{jj'k}^{(m)}$. $\boldsymbol{\eta}_{jj'}$s can then be updated in closed form as
\begin{equation}
\label{closeupdateeta}
    \hat{\boldsymbol{\eta}}=\Big(\tilde{\boldsymbol{X}}^{\top}\tilde{\boldsymbol{X}}+n\lambda_2\boldsymbol{D}^{(m)}\Big)^{-1}\Big(\tilde{\boldsymbol{X}}^{\top}\tilde{\boldsymbol{y}}+\lambda_2 \cdot \boldsymbol{C}^{(m)}\Big).
\end{equation}

\subsection*{\normalfont{A.4.} \textit{Algorithm}}
\label{algorithm}
We describe the full algorithm for estimating $\boldsymbol{\beta}_j$'s and $\boldsymbol{\eta}_{jj'}$'s in (\ref{HiGLASSOcriterion}). We first fix $\boldsymbol{\eta}_{jj'}$ to estimate $\boldsymbol{\beta}_j$, then fix $\boldsymbol{\beta}_j$ to estimate $\boldsymbol{\eta}_{jj'}$, and iterate the two steps until convergence. The algorithm can be summarized as follows:
\begin{enumerate}
\item  Obtain basis-expanded main effect matrices for each covariate, denoted by $\boldsymbol{X}_j$ for $j = 1, \ldots, S$. Normalize $\boldsymbol{X}_j$. Calculate interaction design matrices $\boldsymbol{X}_{jj'}$ from the normalized $\boldsymbol{X}_j$ for $1 \leq j \leq j' \leq S$. Normalize $\boldsymbol{X}_{jj'}$. Orthogonalize $\boldsymbol{X}_j$ and $\boldsymbol{X}_{jj'}$ using QR decomposition and center the response vector $\boldsymbol{y}$. Scale $\boldsymbol{X}_j$ and $\boldsymbol{X}_{jj'}$ to have unit variance.
\item Initialize $\hat{\boldsymbol{\beta}}_j^{(0)}$ for $j = 1, \cdots, S$ and $\hat{\boldsymbol{\eta}}_{jj'}^{(0)}$ for $1 \leq j < j' \leq S$. Set $m = 1$. A feasible choice for the initialization $\hat{\boldsymbol{\beta}}_j^{(0)}$ and $\hat{\boldsymbol{\eta}}_{jj'}^{(0)}$ can be obtained using the adaptive elastic-net estimator. We use this as the initialization in our implementation.
\item For each $j$ in $1, \cdots, S$, update $\hat{\boldsymbol{\beta}}_j^{(m)}$ via closed-form formula in (\ref{closeupdatebeta}), given $\hat{\boldsymbol{\eta}}_{kj}^{(m - 1)}$ and $\hat{\boldsymbol{\beta}}_k^{(m)}$ for $k < j$, and $\hat{\boldsymbol{\eta}}_{jl}^{(m - 1)}$  and $\hat{\boldsymbol{\beta}}_l^{(m - 1)}$ for $l > j$. A backtracking line search algorithm is followed to guarantee that  $\hat{\boldsymbol{\beta}}_j^{(m)}$ leads to a lower value of the objective function (\ref{HiGLASSOcriterionmain}) than $\hat{\boldsymbol{\beta}}_j^{(m)}$.
\item Given $\hat{\boldsymbol{\beta}}_j^{(m)}$ for $j = 1, \cdots, S$, update the $\hat{\boldsymbol{\eta}}_{jj'}^{(m)}$'s via the closed-form formula in (\ref{closeupdateeta}). A backtracking line search algorithm is followed to guarantee that the $\hat{\boldsymbol{\eta}}_{jj'}^{(m)}$'s lead to a lower value of the objective function in (\ref{HiGLASSOcriterionint}) compared to the $\hat{\boldsymbol{\eta}}_{jj'}^{(m - 1)}$'s.
\item Stop if change in the penalized likelihood is less than a pre-specified margin $\delta$, namely
\begin{equation*}
|P_n^{(m-1)}-P_n^{(m)}|<\delta.
\end{equation*}
where $P_n^{(m)}$ is the value of (\ref{HiGLASSOcriterion}) evaluated at the $\hat{\boldsymbol{\beta}}_j^{(m)}$'s and $\hat{\boldsymbol{\eta}}_{jj'}^{(m)}$'s.
\end{enumerate}

\noindent \textbf{Remark 2}: We note that there is no guarantee that each of the $S+1$ updates decreases the value of penalized least squares criterion since we utilize approximations to the original penalty. We therefore employ a backtracking line search algorithm \cite{dennis1996} to ensure that the penalized least squares criterion monotonically decreases throughout the entire procedure. The maximum amount to move along a given search direction is determined by the Armijo-Goldstein condition \cite{armijo1966}.\\

\noindent \textbf{Remark 3}: Steps (3) and (4) in the HiGLASSO algorithm could be easily modified to accommodate objective functions without the least squares criterion. However, closed-form updates may not be avilable, thus requiring one-step gradient descent.

\newpage

\section*{Web Appendix B: Sparsistency proof details}

\subsection*{\normalfont{B.1.} \textit{Notation}}
\label{notation}

\noindent Let $\boldsymbol{X}=[\boldsymbol{X}_1, \cdots,\boldsymbol{X}_S, \boldsymbol{X}_{12}, \cdots,  \boldsymbol{X}_{S, S-1}]$ be the design matrix containing main effect and interaction terms. Without loss of generality, we rearrange the group indices so that the first $s_0 \leq S$ groups of predictors have nonzero main effects. Suppose there are $i_0$ nonzero two-way interaction terms out of at most $s_0(s_0 - 1)/ 2$ possible pairs under strong heredity constraints.

The HiGLASSO estimator is defined as:
\begin{align}
 & \underset{\boldsymbol{\beta}_j, \boldsymbol{\eta}_{jj'}}{\arg\min} \hspace{1 mm} \frac{1}{2}\Big\Vert\boldsymbol{y}-\sum_{j=1}^{S}\boldsymbol{X}_j\boldsymbol{\beta}_j-\sum_{1 \leq j<j' \leq S}\boldsymbol{X}_{jj'}[\boldsymbol{\eta}_{jj'} \odot (\boldsymbol{\beta}_j \otimes \boldsymbol{\beta}_{j'})]\Big\Vert_2^2 \nonumber & \label{HiGLASSOcriterionfull} \\
  & + \lambda_1(n)\sum_{j = 1}^{S}w_j(\boldsymbol{\beta}_j)||\boldsymbol{\beta}_j||_2 + \lambda_2(n)\sum_{1 \leq j<j' \leq S}w_{jj'}(\boldsymbol{\eta}_{jj'})||\boldsymbol{\eta}_{jj'}||_2. &
\end{align}

\subsection*{\normalfont{B.2.} \textit{Directional Derivatives of HiGLASSO Objective Function}}
\label{direcderiv}

\noindent Consider the following function \begin{align}
  f(\boldsymbol{\beta}_1,...,\boldsymbol{\beta}_S,\boldsymbol{\eta}_{12},...,\boldsymbol{\eta}_{S-1,S}) = \frac{1}{2}\Big\Vert\boldsymbol{y}-\sum_{j=1}^{S}\boldsymbol{X}_j\boldsymbol{\beta}_j-\sum_{1 \leq j<j' \leq S}\boldsymbol{X}_{jj'}[\boldsymbol{\eta}_{jj'} \odot (\boldsymbol{\beta}_j \otimes \boldsymbol{\beta}_{j'})]\Big\Vert_2^2 \nonumber
\end{align}

First we will calculate the directional derivative in the $\boldsymbol{u}$ direction with respect to $\boldsymbol{\beta}_k$. By definition the directional derivative is given by: $$\lim_{t \to 0^+} \frac{f(\boldsymbol{\beta}_1,...,\boldsymbol{\beta}_{k-1},\boldsymbol{\beta}_k+t\boldsymbol{u},\boldsymbol{\beta}_{k+1},...,\boldsymbol{\beta}_S,\boldsymbol{\eta}_{12},...,\boldsymbol{\eta}_{S-1,S})-f(\boldsymbol{\beta}_1,...,\boldsymbol{\beta}_S,\boldsymbol{\eta}_{12},...,\boldsymbol{\eta}_{S-1,S})}{t}$$

$$f(\boldsymbol{\beta}_1,...,\boldsymbol{\beta}_{k-1},\boldsymbol{\beta}_k+t\boldsymbol{u},\boldsymbol{\beta}_{k+1},...,\boldsymbol{\beta}_S,\boldsymbol{\eta}_{12},...,\boldsymbol{\eta}_{S-1,S})$$ $$= \frac{1}{2}\Big\Vert\boldsymbol{y}-\boldsymbol{X}_k(\boldsymbol{\beta}_k+t\boldsymbol{u})-\sum_{j \neq k}\boldsymbol{X}_j\boldsymbol{\beta}_j-\sum_{1 \leq k<j' \leq S}\boldsymbol{X}_{kj'}[\boldsymbol{\eta}_{kj'} \odot (\boldsymbol{\beta}_k + t\boldsymbol{u} \otimes \boldsymbol{\beta}_{j'})]$$ $$-\sum_{1 \leq j<k \leq S}\boldsymbol{X}_{jk}[\boldsymbol{\eta}_{jk} \odot (\boldsymbol{\beta}_j \otimes \boldsymbol{\beta}_k+t\boldsymbol{u})] - \sum_{1 \leq j<j' \leq S : j,j' \neq k}\boldsymbol{X}_{jj'}[\boldsymbol{\eta}_{jj'} \odot (\boldsymbol{\beta}_j \otimes \boldsymbol{\beta}_{j'})]\Big\Vert_2^2$$

\noindent Note that $$\boldsymbol{\eta}_{jk} \odot (\boldsymbol{\beta}_j \otimes \boldsymbol{\beta}_k + t\boldsymbol{u}) = \boldsymbol{\eta}_{jk} \odot (\boldsymbol{\beta}_j \otimes \boldsymbol{\beta}_k) + t(\boldsymbol{\eta}_{jk} \odot (\boldsymbol{\beta}_j \otimes \boldsymbol{u}))$$ and $$\boldsymbol{\eta}_{kj'} \odot (\boldsymbol{\beta}_k+t\boldsymbol{u} \otimes \boldsymbol{\beta}_{j'}) = \boldsymbol{\eta}_{kj'} \odot (\boldsymbol{\beta}_k \otimes \boldsymbol{\beta}_{j'}) + t(\boldsymbol{\eta}_{kj'} \odot (\boldsymbol{u} \otimes \boldsymbol{\beta}_{j'}))$$

\noindent Thus, the expression becomes $$\frac{1}{2}\Big\Vert\boldsymbol{y}-t\boldsymbol{X}_k\boldsymbol{u}-\sum_{j = 1}^{S}\boldsymbol{X}_j\boldsymbol{\beta}_j-t\sum_{1 \leq k<j' \leq S}\boldsymbol{X}_{kj'}[\boldsymbol{\eta}_{kj'} \odot (\boldsymbol{u} \otimes \boldsymbol{\beta}_{j'})]$$ $$-t\sum_{1 \leq j<k \leq S}\boldsymbol{X}_{jk}[\boldsymbol{\eta}_{jk} \odot (\boldsymbol{\beta}_j \otimes \boldsymbol{u})] - \sum_{1 \leq j<j' \leq S}\boldsymbol{X}_{jj'}[\boldsymbol{\eta}_{jj'} \odot (\boldsymbol{\beta}_j \otimes \boldsymbol{\beta}_{j'})]\Big\Vert_2^2$$

\noindent Observe that as we take the limit to $0$ we get that the terms with a $t^2$ term go to $0$ and the terms without a $t$ cancel with $f(\boldsymbol{\beta}_1,...,\boldsymbol{\beta}_S,\boldsymbol{\eta}_{12},...,\boldsymbol{\eta}_{S-1,S})$. Therefore, we only need to keep track of the terms that are linear in $t$. To simplify notation, let $$\boldsymbol{y} - \boldsymbol{X}\boldsymbol{\theta} = \boldsymbol{y} - \sum_{j = 1}^{S}\boldsymbol{X}_j\boldsymbol{\beta}_j - \sum_{1 \leq j<j' \leq S}\boldsymbol{X}_{jj'}[\boldsymbol{\eta}_{jj'} \odot (\boldsymbol{\beta}_j \otimes \boldsymbol{\beta}_{j'})]$$

\noindent Then, the expression becomes $$\frac{1}{2}\Big\Vert \boldsymbol{y} - \boldsymbol{X}\boldsymbol{\theta}-t\boldsymbol{X}_k\boldsymbol{u}-t\sum_{1 \leq k<j' \leq S}\boldsymbol{X}_{kj'}[\boldsymbol{\eta}_{kj'} \odot (\boldsymbol{u} \otimes \boldsymbol{\beta}_{j'})]-t\sum_{1 \leq j<k \leq S}\boldsymbol{X}_{jk}[\boldsymbol{\eta}_{jk} \odot (\boldsymbol{\beta}_j \otimes \boldsymbol{u})]\Big\Vert_2^2$$

\noindent Therefore, the directional derivative is:

$$-\Bigg(\boldsymbol{X}_k\boldsymbol{u}+\sum_{1 \leq k<j' \leq S}\boldsymbol{X}_{kj'}[\boldsymbol{\eta}_{kj'} \odot (\boldsymbol{u} \otimes \boldsymbol{\beta}_{j'})]+\sum_{1 \leq j<k \leq S}\boldsymbol{X}_{jk}[\boldsymbol{\eta}_{jk} \odot (\boldsymbol{\beta}_j \otimes \boldsymbol{u})]\Bigg)^{\top}\Big(\boldsymbol{y} - \boldsymbol{X}\boldsymbol{\theta}\Big)$$

\noindent Lastly, from the proof of Theorem 1 in \cite{pan2016}, we have that the directional derivative of $\lambda_1(n)w_k(\boldsymbol{\beta}_k)\big\Vert\boldsymbol{\beta}_k\big\Vert_2$ in the $\boldsymbol{u}$ direction evaluated at zero is $\lambda_1(n)$.

\vspace{2 mm}

Next we will calculate the directional derivative in the $\boldsymbol{u}$ direction with respect to $\boldsymbol{\eta}_{kk'}$. By definition the directional derivative is given by: $$\lim_{t \to 0^+} \frac{f(\boldsymbol{\beta}_1,...,\boldsymbol{\beta}_S,\boldsymbol{\eta}_{12},...,\boldsymbol{\eta}_{kk'}+t\boldsymbol{u},...,\boldsymbol{\eta}_{S-1,S})-f(\boldsymbol{\beta}_1,...,\boldsymbol{\beta}_S,\boldsymbol{\eta}_{12},...,\boldsymbol{\eta}_{S-1,S})}{t},$$

$$f(\boldsymbol{\beta}_1,...,\boldsymbol{\beta}_S,\boldsymbol{\eta}_{12},...,\boldsymbol{\eta}_{kk'}+t\boldsymbol{u},...,\boldsymbol{\eta}_{S-1,S}) = \frac{1}{2}\Big\Vert\boldsymbol{y}-\sum_{j = 1}^{S}\boldsymbol{X}_j\boldsymbol{\beta}_j-\boldsymbol{X}_{kk'}[(\boldsymbol{\eta}_{kk'} + t\boldsymbol{u}) \odot (\boldsymbol{\beta}_k \otimes \boldsymbol{\beta}_{k'})]$$ $$- \sum_{1 \leq j<j' \leq S : (j,j') \neq (k,k')}\boldsymbol{X}_{jj'}[\boldsymbol{\eta}_{jj'} \odot (\boldsymbol{\beta}_j \otimes \boldsymbol{\beta}_{j'})]\Big\Vert_2^2$$

\noindent Again, note that $$(\boldsymbol{\eta}_{kk'} + t\boldsymbol{u}) \odot (\boldsymbol{\beta}_k \otimes \boldsymbol{\beta}_{k'}) = \boldsymbol{\eta}_{kk'} \odot (\boldsymbol{\beta}_k \otimes \boldsymbol{\beta}_{k'}) + t(\boldsymbol{u} \odot (\boldsymbol{\beta}_k \otimes \boldsymbol{\beta}_{k'}))$$

\noindent Thus the expression becomes

$$\frac{1}{2}\Big\Vert\boldsymbol{y}-\sum_{j = 1}^{S}\boldsymbol{X}_j\boldsymbol{\beta}_j- \sum_{1 \leq j<j' \leq S}\boldsymbol{X}_{jj'}[\boldsymbol{\eta}_{jj'} \odot (\boldsymbol{\beta}_j \otimes \boldsymbol{\beta}_{j'})] - t\boldsymbol{X}_{kk'}[\boldsymbol{u} \odot (\boldsymbol{\beta}_k \otimes \boldsymbol{\beta}_{k'})] \Big\Vert_2^2$$

$$= \frac{1}{2}\Big\Vert \boldsymbol{y} - \boldsymbol{X}\boldsymbol{\theta} - t\boldsymbol{X}_{kk'}[\boldsymbol{u} \odot (\boldsymbol{\beta}_k \otimes \boldsymbol{\beta}_{k'})] \Big\Vert_2^2$$

\noindent Following the same argument as above the directional derivative of $\boldsymbol{\beta}_k$, as we take the limit to $0$ we get that the terms with a $t^2$ term go to $0$ and the terms without a $t$ cancel with $f(\boldsymbol{\beta}_1,...,\boldsymbol{\beta}_S,\boldsymbol{\eta}_{12},...,\boldsymbol{\eta}_{S-1,S})$. Therefore, we only need to keep track of the terms that are linear in $t$. That is, the directional derivative is,

$$-\Big(\boldsymbol{X}_{kk'}[\boldsymbol{u} \odot (\boldsymbol{\beta}_k \otimes \boldsymbol{\beta}_{k'})]\Big)^{\top}\Big(\boldsymbol{y} - \boldsymbol{X}\boldsymbol{\theta}\Big)$$

\noindent Again, from the proof of Theorem 1 in \cite{pan2016}, we have that the directional derivative of $\lambda_2(n)w_{kk'}(\boldsymbol{\eta}_{kk'})\big\Vert\boldsymbol{\eta}_{kk'}\big\Vert_2$ in the $\boldsymbol{u}$ direction evaluated at zero is $\lambda_2(n)$.

\subsection*{\normalfont{B.3.} \textit{Derivative of HiGLASSO Objective Function}}
\label{gradderiv}

First we calculate the derivative with respect to $\boldsymbol{\beta}_k$:

$$\frac{\partial}{\partial\boldsymbol{\beta}_k}f(\boldsymbol{\beta}_1,...,\boldsymbol{\beta}_S,\boldsymbol{\eta}_{12},...,\boldsymbol{\eta}_{S-1,S}) = \frac{\partial}{\partial\boldsymbol{\beta}_k}\Bigg[\frac{1}{2}\Big\Vert\boldsymbol{y}-\sum_{j=1}^{S}\boldsymbol{X}_j\boldsymbol{\beta}_j-\sum_{1 \leq j<j' \leq S}\boldsymbol{X}_{jj'}[\boldsymbol{\eta}_{jj'} \odot (\boldsymbol{\beta}_j \otimes \boldsymbol{\beta}_{j'})]\Big\Vert_2^2\Bigg]$$

$$= \Bigg(\frac{\partial}{\partial\boldsymbol{\beta}_k}\Bigg[\boldsymbol{y}-\sum_{j=1}^{S}\boldsymbol{X}_j\boldsymbol{\beta}_j-\sum_{1 \leq j<j' \leq S}\boldsymbol{X}_{jj'}[\boldsymbol{\eta}_{jj'} \odot (\boldsymbol{\beta}_j \otimes \boldsymbol{\beta}_{j'})]\Bigg]\Bigg)^{\top}\Big(\boldsymbol{y} - \boldsymbol{X}\boldsymbol{\theta}\Big)$$

$$= \Bigg(\frac{\partial}{\partial\boldsymbol{\beta}_k}\Bigg[-\boldsymbol{X}_k\boldsymbol{\beta}_k-\sum_{1 \leq k<j' \leq S}\boldsymbol{X}_{kj'}[\boldsymbol{\eta}_{kj'} \odot (\boldsymbol{\beta}_k \otimes \boldsymbol{\beta}_{j'})]-\sum_{1 \leq j<k \leq S}\boldsymbol{X}_{jk}[\boldsymbol{\eta}_{jk} \odot (\boldsymbol{\beta}_j \otimes \boldsymbol{\beta}_{k})]\Bigg]\Bigg)^{\top}\Big(\boldsymbol{y} - \boldsymbol{X}\boldsymbol{\theta}\Big)$$

$$= -\Bigg[\boldsymbol{X}_k+\sum_{1 \leq k<j' \leq S}\boldsymbol{X}_{kj'}\frac{\partial}{\partial\boldsymbol{\beta}_k}[\boldsymbol{\eta}_{kj'} \odot (\boldsymbol{\beta}_k \otimes \boldsymbol{\beta}_{j'})]+\sum_{1 \leq j<k \leq S}\boldsymbol{X}_{jk}\frac{\partial}{\partial\boldsymbol{\beta}_k}[\boldsymbol{\eta}_{jk} \odot (\boldsymbol{\beta}_j \otimes \boldsymbol{\beta}_{k})]\Bigg]^{\top}\Big(\boldsymbol{y} - \boldsymbol{X}\boldsymbol{\theta}\Big)$$

$$= -\Bigg[\boldsymbol{X}_k+\sum_{1 \leq k<j' \leq S}\boldsymbol{X}_{kj'}\bigg[\text{diag}\big(\boldsymbol{\eta}_{kj'}\big) \frac{\partial}{\partial\boldsymbol{\beta}_k}(\boldsymbol{\beta}_k \otimes \boldsymbol{\beta}_{j'})\bigg]+\sum_{1 \leq j<k \leq S}\boldsymbol{X}_{jk}\bigg[\text{diag}\big(\boldsymbol{\eta}_{jk}\big) \frac{\partial}{\partial\boldsymbol{\beta}_k}(\boldsymbol{\beta}_j \otimes \boldsymbol{\beta}_{k})\bigg]\Bigg]^{\top}\Big(\boldsymbol{y} - \boldsymbol{X}\boldsymbol{\theta}\Big)$$

$$= -\bigg[\boldsymbol{X}_k+\sum_{1 \leq k<j' \leq S}\boldsymbol{X}_{kj'}\Big[\text{diag}\big(\boldsymbol{\eta}_{kj'}\big) (\boldsymbol{I} \otimes \boldsymbol{\beta}_{j'})\Big]+\sum_{1 \leq j<k \leq S}\boldsymbol{X}_{jk}\Big[\text{diag}\big(\boldsymbol{\eta}_{jk}\big) (\boldsymbol{\beta}_j \otimes \boldsymbol{I})\Big]\bigg]^{\top}\Big(\boldsymbol{y} - \boldsymbol{X}\boldsymbol{\theta}\Big)$$

\noindent The derivative of the penalty function is:

$$\frac{\partial}{\partial\boldsymbol{\beta}_k}w_k(\boldsymbol{\beta}_k)||\boldsymbol{\beta}_k||_2 = \frac{\partial}{\partial\boldsymbol{\beta}_k}\exp\bigg(-\frac{||\boldsymbol{\beta}_k||_{\infty}}{\sigma(n)}\bigg)||\boldsymbol{\beta}_k||_2$$

$$= ||\boldsymbol{\beta}_k||_2\frac{\partial}{\partial\boldsymbol{\beta}_k}\exp\bigg(-\frac{||\boldsymbol{\beta}_k||_{\infty}}{\sigma(n)}\bigg) + \exp\bigg(-\frac{||\boldsymbol{\beta}_k||_{\infty}}{\sigma(n)}\bigg)\frac{\partial}{\partial\boldsymbol{\beta}_k}||\boldsymbol{\beta}_k||_2$$

$$= ||\boldsymbol{\beta}_k||_2\Bigg(-\frac{1}{\sigma(n)}\exp\bigg(-\frac{||\boldsymbol{\beta}_k||_{\infty}}{\sigma(n)}\bigg)\sum_{l=1}^{p_k}\text{sign}(\beta_{kl})\Vec{\boldsymbol{e}}_lI\Big(\beta_{kl}=||\boldsymbol{\beta}_k||_{\infty}\Big)\Bigg)+ \exp\bigg(-\frac{||\boldsymbol{\beta}_k||_{\infty}}{\sigma(n)}\bigg)\Big(||\boldsymbol{\beta}_k||_2\Big)^{-1}\boldsymbol{\beta}_k,$$

\noindent where $\Vec{\boldsymbol{e}}_l$ is the standard basis vector of dimension $p_k$ such that the $l$-th component is equal to $1$.

\vspace{2 mm}

\noindent Next we calculate the derivative with respect to $\boldsymbol{\eta}_{kk'}$:

$$\frac{\partial}{\partial\boldsymbol{\eta}_{kk'}}f(\boldsymbol{\beta}_1,...,\boldsymbol{\beta}_S,\boldsymbol{\eta}_{12},...,\boldsymbol{\eta}_{S-1,S})$$

$$= \frac{\partial}{\partial\boldsymbol{\eta}_{kk'}}\Bigg[\frac{1}{2}\Big\Vert\boldsymbol{y}-\sum_{j=1}^{S}\boldsymbol{X}_j\boldsymbol{\beta}_j-\sum_{1 \leq j<j' \leq S}\boldsymbol{X}_{jj'}[\boldsymbol{\eta}_{jj'} \odot (\boldsymbol{\beta}_j \otimes \boldsymbol{\beta}_{j'})]\Big\Vert_2^2\Bigg]$$

$$= \Bigg(\frac{\partial}{\partial\boldsymbol{\eta}_{kk'}}\Bigg[\boldsymbol{y}-\sum_{j=1}^{S}\boldsymbol{X}_j\boldsymbol{\beta}_j-\sum_{1 \leq j<j' \leq S}\boldsymbol{X}_{jj'}[\boldsymbol{\eta}_{jj'} \odot (\boldsymbol{\beta}_j \otimes \boldsymbol{\beta}_{j'})]\Bigg]\Bigg)^{\top}\Big(\boldsymbol{y} - \boldsymbol{X}\boldsymbol{\theta}\Big)$$

$$= -\Bigg(\frac{\partial}{\partial\boldsymbol{\eta}_{kk'}}\Big[\boldsymbol{X}_{kk'}[\boldsymbol{\eta}_{kk'} \odot (\boldsymbol{\beta}_k \otimes \boldsymbol{\beta}_{k'})]\Big]\Bigg)^{\top}\Big(\boldsymbol{y} - \boldsymbol{X}\boldsymbol{\theta}\Big)$$

$$= -\Bigg[\boldsymbol{X}_{kk'}\bigg[\frac{\partial}{\partial\boldsymbol{\eta}_{kk'}}\boldsymbol{\eta}_{kk'} \odot \text{diag}(\boldsymbol{\beta}_k \otimes \boldsymbol{\beta}_{k'})\bigg]\Bigg]^{\top}\Big(\boldsymbol{y} - \boldsymbol{X}\boldsymbol{\theta}\Big)$$

$$= -\Bigg(\boldsymbol{X}_{kk'}\bigg[\boldsymbol{I} \odot \text{diag}(\boldsymbol{\beta}_k \otimes \boldsymbol{\beta}_{k'})\bigg]\Bigg)^{\top}\Big(\boldsymbol{y} - \boldsymbol{X}\boldsymbol{\theta}\Big)$$

$$= -\bigg(\boldsymbol{X}_{kk'}\Big[\text{diag}(\boldsymbol{\beta}_k \otimes \boldsymbol{\beta}_{k'})\Big]\bigg)^{\top}\Big(\boldsymbol{y} - \boldsymbol{X}\boldsymbol{\theta}\Big)$$

\noindent The derivative of the penalty function is:

$$\frac{\partial}{\partial\boldsymbol{\eta}_{kk'}}w_{kk'}(\boldsymbol{\eta}_{kk'})||\boldsymbol{\eta}_{kk'}||_2 = \frac{\partial}{\partial\boldsymbol{\eta}_{kk'}}\exp\bigg(-\frac{||\boldsymbol{\eta}_{kk'}||_{\infty}}{\sigma(n)}\bigg)||\boldsymbol{\eta}_{kk'}||_2$$

$$= ||\boldsymbol{\eta}_{kk'}||_2\frac{\partial}{\partial\boldsymbol{\eta}_{kk'}}\exp\bigg(-\frac{||\boldsymbol{\eta}_{kk'}||_{\infty}}{\sigma(n)}\bigg) + \exp\bigg(-\frac{||\boldsymbol{\eta}_{kk'}||_{\infty}}{\sigma(n)}\bigg)\frac{\partial}{\partial\boldsymbol{\eta}_{kk'}}||\boldsymbol{\eta}_{kk'}||_2$$

$$= ||\boldsymbol{\eta}_{kk'}||_2\Bigg(-\frac{1}{\sigma(n)}\exp\bigg(-\frac{||\boldsymbol{\eta}_{kk'}||_{\infty}}{\sigma(n)}\bigg)\sum_{l=1}^{p_kp_{k'}}\text{sign}(\eta_{kk'l})\Vec{\boldsymbol{e}}_lI\Big(\eta_{kk'l}=||\boldsymbol{\eta}_{kk'}||_{\infty}\Big)\Bigg) + \exp\bigg(-\frac{||\boldsymbol{\eta}_{kk'}||_{\infty}}{\sigma(n)}\bigg)\Big(||\boldsymbol{\eta}_{kk'}||_2\Big)^{-1}\boldsymbol{\eta}_{kk'},$$

\noindent where $\Vec{\boldsymbol{e}}_l$ is the standard basis vector of dimension $p_kp_{k'}$ such that the $l$-th component is equal to $1$.

\subsection*{\normalfont{B.4.} \textit{Sparsistency Proof}}
\label{sparseproof}

The proof closely follows the proof of Theorem 1 in \cite{pan2016}. Define the HiGLASSO estimator of a re-parameterized version of (\ref{HiGLASSOcriterionfull}) such that only the covariates corresponding to the non-zero coefficient set are included:

\begin{equation}
    \underset{\boldsymbol{\theta}_{\mathcal{P}}}{\arg\min} \bigg\{||\boldsymbol{y}-\boldsymbol{X}_{\mathcal{P}}\boldsymbol{\theta}_{\mathcal{P}}||_2^2 + \lambda_{1}(n)\sum_{j \in \mathcal{P}_1}w_j(\boldsymbol{\theta}_j)||\boldsymbol{\theta}_j||_2 + \lambda_{2}(n)\sum_{(j,j') \in \mathcal{P}_2}w_{jj'}(\boldsymbol{\eta}_{jj'})||\boldsymbol{\eta}_{jj'}||_2\bigg\} \label{HiGLASSOcriterionsub}.
\end{equation}

\noindent Let $\tilde{\boldsymbol{\theta}}_{\mathcal{P}}$ be the solution to (\ref{HiGLASSOcriterionsub}). From the assumptions of the Theorem we have that

\begin{align*}
\frac{1}{n}\boldsymbol{X}^{\top}\boldsymbol{y} & = \frac{1}{n}\boldsymbol{X}^{\top}\boldsymbol{X}_{\mathcal{P}}\boldsymbol{\theta}_{\mathcal{P}} + \frac{1}{n}\boldsymbol{X}^{\top}\boldsymbol{\epsilon} \nonumber & \\
& = \bigg[E\bigg(\frac{1}{n}\boldsymbol{X}^{\top}\boldsymbol{X}_{\mathcal{P}}\bigg)+O_p\big(n^{-1/2}\big)\bigg]\boldsymbol{\theta}_{\mathcal{P}} + O_p\big(n^{-1/2}\big) \nonumber &\\
& = E\bigg(\frac{1}{n}\boldsymbol{X}^{\top}\boldsymbol{X}_{\mathcal{P}}\bigg)\boldsymbol{\theta}_{\mathcal{P}} + O_p\big(n^{-1/2}\big) \nonumber, &
\end{align*}
which implies that
\begin{equation}
\frac{1}{n}\boldsymbol{X}^{\top}\boldsymbol{y}-\frac{1}{n}\boldsymbol{X}^{\top}\boldsymbol{X}_{\mathcal{P}}\tilde{\boldsymbol{\theta}}_{\mathcal{P}} =E\bigg(\frac{1}{n}\boldsymbol{X}^{\top}\boldsymbol{X}_{\mathcal{P}}\bigg)(\boldsymbol{\theta}_{\mathcal{P}}-\tilde{\boldsymbol{\theta}}_{\mathcal{P}})+O_p\big(n^{-1/2}\big) \label{p0}. \\
\end{equation}
(\ref{p0}) can be decomposed as

\begin{equation}
    \frac{1}{n}\boldsymbol{X}_{\mathcal{P}}^{\top}\boldsymbol{y}-\frac{1}{n}\boldsymbol{X}_{\mathcal{P}}^{\top}\boldsymbol{X}_{\mathcal{P}}\tilde{\boldsymbol{\theta}}_{\mathcal{P}} =E\bigg(\frac{1}{n}\boldsymbol{X}_{\mathcal{P}}^{\top}\boldsymbol{X}_{\mathcal{P}}\bigg)(\boldsymbol{\theta}_{\mathcal{P}}-\tilde{\boldsymbol{\theta}}_{\mathcal{P}})+O_p\big(n^{-1/2}\big) \label{p1}
\end{equation}

\begin{equation}
    \frac{1}{n}\boldsymbol{X}_{\mathcal{P}^{\mathsf{c}}}^{\top}\boldsymbol{y}-\frac{1}{n}\boldsymbol{X}_{\mathcal{P}^{\mathsf{c}}}^{\top}\boldsymbol{X}_{\mathcal{P}}\tilde{\boldsymbol{\theta}}_{\mathcal{P}} =E\bigg(\frac{1}{n}\boldsymbol{X}_{\mathcal{P}^{\mathsf{c}}}^{\top}\boldsymbol{X}_{\mathcal{P}}\bigg)(\boldsymbol{\theta}_{\mathcal{P}}-\tilde{\boldsymbol{\theta}}_{\mathcal{P}})+O_p\big(n^{-1/2}\big) \label{p2}
\end{equation}

\noindent From (\ref{p1}) we get
$$\boldsymbol{\theta}_{\mathcal{P}}-\tilde{\boldsymbol{\theta}}_{\mathcal{P}} = E^{-1}\bigg(\frac{1}{n}\boldsymbol{X}_{\mathcal{P}}^{\top}\boldsymbol{X}_{\mathcal{P}}\bigg)\frac{1}{n}\boldsymbol{X}_{\mathcal{P}}^{\top}\big(\boldsymbol{y}-\boldsymbol{X}_{\mathcal{P}}\tilde{\boldsymbol{\theta}}_{\mathcal{P}}\big) + O_p\big(n^{-1/2}\big)$$

\noindent and substituting into (\ref{p2}) we obtain
$$\frac{1}{n}\boldsymbol{X}_{\mathcal{P}^{\mathsf{c}}}^{\top}\big(\boldsymbol{y}-\boldsymbol{X}_{\mathcal{P}}\tilde{\boldsymbol{\theta}}_{\mathcal{P}}\big)=E\bigg(\frac{1}{n}\boldsymbol{X}_{\mathcal{P}^{\mathsf{c}}}^{\top}\boldsymbol{X}_{\mathcal{P}}\bigg)E^{-1}\bigg(\frac{1}{n}\boldsymbol{X}_{\mathcal{P}}^{\top}\boldsymbol{X}_{\mathcal{P}}\bigg)\frac{1}{n}\boldsymbol{X}_{\mathcal{P}}^{\top}\big(\boldsymbol{y}-\boldsymbol{X}_{\mathcal{P}}\tilde{\boldsymbol{\theta}}_{\mathcal{P}}\big) + O_p\big(n^{-1/2}\big).$$

\noindent Multiplying both sides by $n/a_n$, we get
$$\frac{n}{a_n}\bigg(\frac{1}{n}\boldsymbol{X}_{\mathcal{P}^{\mathsf{c}}}^{\top}\big(\boldsymbol{y}-\boldsymbol{X}_{\mathcal{P}}\tilde{\boldsymbol{\theta}}_{\mathcal{P}}\big)\bigg)=E\bigg(\frac{1}{n}\boldsymbol{X}_{\mathcal{P}^{\mathsf{c}}}^{\top}\boldsymbol{X}_{\mathcal{P}}\bigg)E^{-1}\bigg(\frac{1}{n}\boldsymbol{X}_{\mathcal{P}}^{\top}\boldsymbol{X}_{\mathcal{P}}\bigg)\frac{1}{a_n}\boldsymbol{X}_{\mathcal{P}}^{\top}\big(\boldsymbol{y}-\boldsymbol{X}_{\mathcal{P}}\tilde{\boldsymbol{\theta}}_{\mathcal{P}}\big) + O_p\bigg(\frac{\sqrt{n}}{a_n}\bigg).$$

\noindent Therefore, when $b_n \to 0$, $a_n/\sqrt{n} \to \infty$, and $a_n/n \to 0$ we have $$\frac{n}{a_n}\bigg\Vert\frac{1}{n}\boldsymbol{X}_{\mathcal{P}^{\mathsf{c}}}^{\top}\big(\boldsymbol{y}-\boldsymbol{X}_{\mathcal{P}}\tilde{\boldsymbol{\theta}}_{\mathcal{P}}\big)\bigg\Vert_2 \to_p 0,$$ which also implies that $$\frac{n}{\lambda_1(n)}\bigg\Vert\frac{1}{n}\boldsymbol{X}_{[k]}^{\top}\big(\boldsymbol{y}-\boldsymbol{X}_{\mathcal{P}}\tilde{\boldsymbol{\theta}}_{\mathcal{P}}\big)\bigg\Vert_2 \to_p 0, \hspace{2 mm} \forall k \in \mathcal{P}_1^{\mathsf{c}}$$ $$\frac{n}{\lambda_2(n)}\bigg\Vert\frac{1}{n}\boldsymbol{X}_{kk'}^{\top}\big(\boldsymbol{y}-\boldsymbol{X}_{\mathcal{P}}\tilde{\boldsymbol{\theta}}_{\mathcal{P}}\big)\bigg\Vert_2 \to_p 0, \hspace{2 mm} \forall (k,k') \in \mathcal{P}_2^{\mathsf{c}}$$ where $$\boldsymbol{X}_{[k]} = \begin{pmatrix} \boldsymbol{X}_k, & \boldsymbol{X}_{k,k+1}, & \cdots & \boldsymbol{X}_{k,S}, & \boldsymbol{X}_{1,k}, & \cdots & \boldsymbol{X}_{k-1,k} \end{pmatrix}$$ is the submatrix of the design matrix corresponding to the $k$th covariate. These two convergence in probability statements imply that

\begin{equation}
    P\Bigg(\forall k \in \mathcal{P}_1^{\mathsf{c}}, \frac{nB_1}{\lambda_1(n)}\bigg\Vert\frac{1}{n}\boldsymbol{X}_{[k]}^{\top}\big(\boldsymbol{y}-\boldsymbol{X}_{\mathcal{P}}\tilde{\boldsymbol{\theta}}_{\mathcal{P}}\big)\bigg\Vert_2 \leq 1\Bigg) \to 1 \label{conv1}
\end{equation}

\begin{equation}
    P\Bigg(\forall (k,k') \in \mathcal{P}_2^{\mathsf{c}}, \frac{nB_2}{\lambda_2(n)}\bigg\Vert\frac{1}{n}\boldsymbol{X}_{kk'}^{\top}\big(\boldsymbol{y}-\boldsymbol{X}_{\mathcal{P}}\tilde{\boldsymbol{\theta}}_{\mathcal{P}}\big)\bigg\Vert_2 \leq 1\Bigg) \to 1 \label{conv2}
\end{equation}

\noindent for any finite constants $B_1$ and $B_2$.

\vspace{1 mm}

Define $\tilde{\boldsymbol{\theta}}$ as the concatenation of $\tilde{\boldsymbol{\theta}}_{\mathcal{P}_1}$, a vector of zeros with length equal to the number of columns in $\boldsymbol{X}$ corresponding to $\mathcal{P}_1^{\mathsf{c}}$, $\tilde{\boldsymbol{\theta}}_{\mathcal{P}_2}$, and a vector of zeros with length equal to the number of columns in $\boldsymbol{X}$ corresponding to $\mathcal{P}_2^{\mathsf{c}}$. The assumption that the $L_2$ norm of the HiGLASSO estimator is uniformly bounded for all $n$ coupled with (\ref{conv1}) and (\ref{conv2}) imply that with probability approaching one

\begin{equation}
    \frac{1}{n}\tilde{\boldsymbol{C}}_{[k]}^{\top}\boldsymbol{X}_{[k]}^{\top}\big(\boldsymbol{y}-\boldsymbol{X}\tilde{\boldsymbol{\theta}}\big) = \frac{\lambda_1(n)}{n}D_k(\tilde{\boldsymbol{\beta}}_k), \hspace{2 mm} \forall k \in \mathcal{P}_1 \label{kktmain1}
\end{equation}

\begin{equation}
    \frac{1}{n}\text{diag}(\tilde{\boldsymbol{\beta}}_k \otimes \tilde{\boldsymbol{\beta}}_{k'})\boldsymbol{X}_{kk'}^{\top}\big(\boldsymbol{y}-\boldsymbol{X}\tilde{\boldsymbol{\theta}}\big) = \frac{\lambda_2(n)}{n}D_{kk'}(\tilde{\boldsymbol{\eta}}_{kk'}), \hspace{2 mm} \forall (k,k') \in \mathcal{P}_2 \label{kktint1}
\end{equation}

\begin{equation}
    \bigg\Vert\frac{1}{n}\tilde{\boldsymbol{C}}_{[k]}^{\top}\boldsymbol{X}_{[k]}^{\top}\big(\boldsymbol{y}-\boldsymbol{X}\tilde{\boldsymbol{\theta}}\big)\bigg\Vert_2 \leq \frac{\lambda_1(n)}{n}, \hspace{2 mm} \forall k \in \mathcal{P}_1^{\mathsf{c}} \label{kktmain2}
\end{equation}

\begin{equation}
    \bigg\Vert\frac{1}{n}\text{diag}(\tilde{\boldsymbol{\beta}}_k \otimes \tilde{\boldsymbol{\beta}}_{k'})\boldsymbol{X}_{kk'}^{\top}\big(\boldsymbol{y}-\boldsymbol{X}\tilde{\boldsymbol{\theta}}\big)\bigg\Vert_2 \leq \frac{\lambda_2(n)}{n}, \hspace{2 mm} \forall (k,k') \in \mathcal{P}_2^{\mathsf{c}} \label{kktint2}
\end{equation}

\noindent where $$\tilde{\boldsymbol{C}}_{[k]} = \begin{pmatrix} \boldsymbol{I}_{p_k \times p_k} \\ \text{diag}(\tilde{\boldsymbol{\eta}}_{k,k+1}) (\boldsymbol{I}_{p_k \times p_k} \otimes \tilde{\boldsymbol{\beta}}_{k+1}) \\ \vdots \\ \text{diag}(\tilde{\boldsymbol{\eta}}_{k,S}) (\boldsymbol{I}_{p_k \times p_k} \otimes \tilde{\boldsymbol{\beta}}_{S}) \\ \text{diag}(\tilde{\boldsymbol{\eta}}_{1,k}) (\tilde{\boldsymbol{\beta}}_{1} \otimes \boldsymbol{I}_{p_k \times p_k}) \\ \vdots \\ \text{diag}(\tilde{\boldsymbol{\eta}}_{k-1,k}) (\tilde{\boldsymbol{\beta}}_{k-1} \otimes \boldsymbol{I}_{p_k \times p_k}) \end{pmatrix}$$

\vspace{2 mm}

$$D_k(\tilde{\boldsymbol{\beta}}_k) = \frac{\partial}{\partial\boldsymbol{\beta}_k}w_k(\boldsymbol{\beta}_k)||\boldsymbol{\beta}_k||_2\bigg|_{\boldsymbol{\beta_k} = \tilde{\boldsymbol{\beta}}_k}$$

\vspace{2 mm}

$$D_{kk'}(\tilde{\boldsymbol{\eta}}_{kk'}) = \frac{\partial}{\partial\boldsymbol{\eta}_{kk'}}w_{kk'}(\boldsymbol{\eta}_{kk'})||\boldsymbol{\eta}_{kk'}||_2\bigg|_{\boldsymbol{\eta_{kk'}} = \tilde{\boldsymbol{\eta}}_{kk'}}$$

\noindent The directional derivative with respect to $\boldsymbol{\beta}_{k}$ in the $u$ direction of (\ref{HiGLASSOcriterionfull}) is

$$-\Bigg(\boldsymbol{X}_k\boldsymbol{u}+\sum_{1 \leq k<j' \leq S}\boldsymbol{X}_{kj'}[\boldsymbol{\eta}_{kj'} \odot (\boldsymbol{u} \otimes \boldsymbol{\beta}_{j'})]+\sum_{1 \leq j<k \leq S}\boldsymbol{X}_{jk}[\boldsymbol{\eta}_{jk} \odot (\boldsymbol{\beta}_j \otimes \boldsymbol{u})]\Bigg)^{\top}\big(\boldsymbol{y}-\boldsymbol{X}\tilde{\boldsymbol{\theta}}\big) + \lambda_1(n)$$

$$= -\begin{pmatrix} \boldsymbol{u} \\ \boldsymbol{\eta}_{k,k+1} \odot (\boldsymbol{u} \otimes \boldsymbol{\beta}_{k+1}) \\ \vdots \\ \boldsymbol{\eta}_{k,S} \odot (\boldsymbol{u} \otimes \boldsymbol{\beta}_{S}) \\ \boldsymbol{\eta}_{1,k} \odot (\boldsymbol{\beta}_{1} \otimes \boldsymbol{u}) \\ \vdots \\ \boldsymbol{\eta}_{k-1,k} \odot (\boldsymbol{\beta}_{k-1} \otimes \boldsymbol{u}) \end{pmatrix}^{\top}\boldsymbol{X}_{[k]}^{\top}\big(\boldsymbol{y}-\boldsymbol{X}\tilde{\boldsymbol{\theta}}\big) + \lambda_1(n).$$

\noindent For $\tilde{\boldsymbol{\beta}}_j$ and $\tilde{\boldsymbol{\eta}}_{jj'}$ to be the minimizer's of (\ref{HiGLASSOcriterionfull}), we need $$-\begin{pmatrix} \boldsymbol{u} \\ \tilde{\boldsymbol{\eta}}_{k,k+1} \odot (\boldsymbol{u} \otimes \tilde{\boldsymbol{\beta}}_{k+1}) \\ \vdots \\ \tilde{\boldsymbol{\eta}}_{k,S} \odot (\boldsymbol{u} \otimes \tilde{\boldsymbol{\beta}}_{S}) \\ \tilde{\boldsymbol{\eta}}_{1,k} \odot (\tilde{\boldsymbol{\beta}}_{1} \otimes \boldsymbol{u}) \\ \vdots \\ \tilde{\boldsymbol{\eta}}_{k-1,k} \odot (\tilde{\boldsymbol{\beta}}_{k-1} \otimes \boldsymbol{u}) \end{pmatrix}^{\top}\boldsymbol{X}_{[k]}^{\top}\big(\boldsymbol{y}-\boldsymbol{X}\tilde{\boldsymbol{\theta}}\big) + \lambda_1(n) \geq 0,$$

\noindent for all $p_k$ dimensional unit vectors $\boldsymbol{u}$. To verify this we must substitute the negative normalized gradient in for $\boldsymbol{u}$, and see when the inequality holds. The negative normalized gradient is given by $$\boldsymbol{u}^* = \frac{\big(\boldsymbol{X}_{[k]}\boldsymbol{C}_{[k]}\big)^{\top}\big(\boldsymbol{y}-\boldsymbol{X}\tilde{\boldsymbol{\theta}}\big)}{\Big\Vert\big(\boldsymbol{X}_{[k]}\boldsymbol{C}_{[k]}\big)^{\top}\big(\boldsymbol{y}-\boldsymbol{X}\tilde{\boldsymbol{\theta}}\big)\Big\Vert_2}.$$

\noindent Then we have that $$\boldsymbol{u}^{*{\top}}\boldsymbol{X}_k^{\top}\big(\boldsymbol{y}-\boldsymbol{X}\tilde{\boldsymbol{\theta}}\big) = \frac{\big(\boldsymbol{y}-\boldsymbol{X}\tilde{\boldsymbol{\theta}}\big)^{\top}\boldsymbol{X}_{[k]}\boldsymbol{C}_{[k]}\boldsymbol{I}_{p_k \times p_k}^{\top}\boldsymbol{X}_{k}^{\top}\big(\boldsymbol{y}-\boldsymbol{X}\tilde{\boldsymbol{\theta}}\big)}{\Big\Vert\big(\boldsymbol{X}_{[k]}\boldsymbol{C}_{[k]}\big)^{\top}\big(\boldsymbol{y}-\boldsymbol{X}\tilde{\boldsymbol{\theta}}\big)\Big\Vert_2}$$

$$\Big[\tilde{\boldsymbol{\eta}}_{kj'}^{\top} \odot \big(\boldsymbol{u}^{*\top} \otimes \tilde{\boldsymbol{\beta}}_{j'}^{\top} \big)\Big]\boldsymbol{X}_{kj'}^{\top}\big(\boldsymbol{y}-\boldsymbol{X}\tilde{\boldsymbol{\theta}}\big)= \frac{\Big[\tilde{\boldsymbol{\eta}}_{kj'}^{\top} \odot \Big(\Big(\big(\boldsymbol{y}-\boldsymbol{X}\tilde{\boldsymbol{\theta}}\big)^{\top}\boldsymbol{X}_{[k]}\boldsymbol{C}_{[k]}\Big) \otimes \tilde{\boldsymbol{\beta}}_{j'}^{\top} \Big)\Big]\boldsymbol{X}_{kj'}^{\top}\big(\boldsymbol{y}-\boldsymbol{X}\tilde{\boldsymbol{\theta}}\big)}{\Big\Vert\big(\boldsymbol{X}_{[k]}\boldsymbol{C}_{[k]}\big)^{\top}\big(\boldsymbol{y}-\boldsymbol{X}\tilde{\boldsymbol{\theta}}\big)\Big\Vert_2}$$

$$= \frac{\big(\boldsymbol{y}-\boldsymbol{X}\tilde{\boldsymbol{\theta}}\big)^{\top}\boldsymbol{X}_{[k]}\boldsymbol{C}_{[k]}\Big(\boldsymbol{I}_{p_k \times p_k} \otimes \tilde{\boldsymbol{\beta}}_{j'}^{\top} \Big)\text{diag}\big(\tilde{\boldsymbol{\eta}}_{kj'}\big)\boldsymbol{X}_{kj'}^{\top}\big(\boldsymbol{y}-\boldsymbol{X}\tilde{\boldsymbol{\theta}}\big)}{\Big\Vert\big(\boldsymbol{X}_{[k]}\boldsymbol{C}_{[k]}\big)^{\top}\big(\boldsymbol{y}-\boldsymbol{X}\tilde{\boldsymbol{\theta}}\big)\Big\Vert_2}$$

$$\Big[\tilde{\boldsymbol{\eta}}_{jk}^{\top} \odot \big(\tilde{\boldsymbol{\beta}}_{j}^{\top} \otimes \boldsymbol{u}^{*\top}\big)\Big]\boldsymbol{X}_{jk}^{\top}\big(\boldsymbol{y}-\boldsymbol{X}\tilde{\boldsymbol{\theta}}\big) = \frac{\Big[\tilde{\boldsymbol{\eta}}_{jk}^{\top} \odot \Big(\tilde{\boldsymbol{\beta}}_{j}^{\top} \otimes \Big(\big(\boldsymbol{y}-\boldsymbol{X}\tilde{\boldsymbol{\theta}}\big)^{\top}\boldsymbol{X}_{[k]}\boldsymbol{C}_{[k]}\Big)\Big)\Big]\boldsymbol{X}_{jk}^{\top}\big(\boldsymbol{y}-\boldsymbol{X}\tilde{\boldsymbol{\theta}}\big)}{\Big\Vert\big(\boldsymbol{X}_{[k]}\boldsymbol{C}_{[k]}\big)^{\top}\big(\boldsymbol{y}-\boldsymbol{X}\tilde{\boldsymbol{\theta}}\big)\Big\Vert_2}$$

$$= \frac{\big(\boldsymbol{y}-\boldsymbol{X}\tilde{\boldsymbol{\theta}}\big)^{\top}\boldsymbol{X}_{[k]}\boldsymbol{C}_{[k]} \Big(\tilde{\boldsymbol{\beta}}_{j}^{\top} \otimes \boldsymbol{I}_{p_k \times p_k}\Big)\text{diag}(\tilde{\boldsymbol{\eta}}_{jk})\boldsymbol{X}_{jk}^{\top}\big(\boldsymbol{y}-\boldsymbol{X}\tilde{\boldsymbol{\theta}}\big)}{\Big\Vert\big(\boldsymbol{X}_{[k]}\boldsymbol{C}_{[k]}\big)^{\top}\big(\boldsymbol{y}-\boldsymbol{X}\tilde{\boldsymbol{\theta}}\big)\Big\Vert_2}$$

\noindent Substituting this result in, we get: $$-\begin{pmatrix} \boldsymbol{u}^{*} \\ \tilde{\boldsymbol{\eta}}_{k,k+1} \odot (\boldsymbol{u}^{*} \otimes \tilde{\boldsymbol{\beta}}_{k+1}) \\ \vdots \\ \tilde{\boldsymbol{\eta}}_{k,S} \odot (\boldsymbol{u}^{*} \otimes \tilde{\boldsymbol{\beta}}_{S}) \\ \tilde{\boldsymbol{\eta}}_{1,k} \odot (\tilde{\boldsymbol{\beta}}_{1} \otimes \boldsymbol{u}^{*}) \\ \vdots \\ \tilde{\boldsymbol{\eta}}_{k-1,k} \odot (\tilde{\boldsymbol{\beta}}_{k-1} \otimes \boldsymbol{u}^{*}) \end{pmatrix}^{\top}\boldsymbol{X}_{[k]}^{\top}\big(\boldsymbol{y}-\boldsymbol{X}\tilde{\boldsymbol{\theta}}\big)$$

$$= -\frac{\big(\boldsymbol{y}-\boldsymbol{X}\tilde{\boldsymbol{\theta}}\big)^{\top}\boldsymbol{X}_{[k]}\boldsymbol{C}_{[k]}\boldsymbol{I}_{p_k \times p_k}^{\top}\boldsymbol{X}_{k}^{\top}\big(\boldsymbol{y}-\boldsymbol{X}\tilde{\boldsymbol{\theta}}\big)}{\Big\Vert\big(\boldsymbol{X}_{[k]}\boldsymbol{C}_{[k]}\big)^{\top}\big(\boldsymbol{y}-\boldsymbol{X}\tilde{\boldsymbol{\theta}}\big)\Big\Vert_2}$$

$$- \sum_{j' > k}\Bigg[\frac{\big(\boldsymbol{y}-\boldsymbol{X}\tilde{\boldsymbol{\theta}}\big)^{\top}\boldsymbol{X}_{[k]}\boldsymbol{C}_{[k]}\Big(\boldsymbol{I}_{p_k \times p_k} \otimes \tilde{\boldsymbol{\beta}}_{j'}^{\top} \Big)\text{diag}\big(\tilde{\boldsymbol{\eta}}_{kj'}\big)\boldsymbol{X}_{kj'}^{\top}\big(\boldsymbol{y}-\boldsymbol{X}\tilde{\boldsymbol{\theta}}\big)}{\Big\Vert\big(\boldsymbol{X}_{[k]}\boldsymbol{C}_{[k]}\big)^{\top}\big(\boldsymbol{y}-\boldsymbol{X}\tilde{\boldsymbol{\theta}}\big)\Big\Vert_2}\Bigg]$$

$$- \sum_{j < k}\Bigg[\frac{\big(\boldsymbol{y}-\boldsymbol{X}\tilde{\boldsymbol{\theta}}\big)^{\top}\boldsymbol{X}_{[k]}\boldsymbol{C}_{[k]} \Big(\tilde{\boldsymbol{\beta}}_{j}^{\top} \otimes \boldsymbol{I}_{p_k \times p_k}\Big)\text{diag}(\tilde{\boldsymbol{\eta}}_{jk})\boldsymbol{X}_{jk}^{\top}\big(\boldsymbol{y}-\boldsymbol{X}\tilde{\boldsymbol{\theta}}\big)}{\Big\Vert\big(\boldsymbol{X}_{[k]}\boldsymbol{C}_{[k]}\big)^{\top}\big(\boldsymbol{y}-\boldsymbol{X}\tilde{\boldsymbol{\theta}}\big)\Big\Vert_2}\Bigg]$$

$$= -\frac{\big(\boldsymbol{y}-\boldsymbol{X}\tilde{\boldsymbol{\theta}}\big)^{\top}\boldsymbol{X}_{[k]}\boldsymbol{C}_{[k]}\boldsymbol{C}_{[k]}^{\top}\boldsymbol{X}_{[k]}^{\top}\big(\boldsymbol{y}-\boldsymbol{X}\tilde{\boldsymbol{\theta}}\big)}{\Big\Vert\big(\boldsymbol{X}_{[k]}\boldsymbol{C}_{[k]}\big)^{\top}\big(\boldsymbol{y}-\boldsymbol{X}\tilde{\boldsymbol{\theta}}\big)\Big\Vert_2}$$

$$= -\frac{\Big\Vert\big(\boldsymbol{X}_{[k]}\boldsymbol{C}_{[k]}\big)^{\top}\big(\boldsymbol{y}-\boldsymbol{X}\tilde{\boldsymbol{\theta}}\big)\Big\Vert_2^2}{\Big\Vert\big(\boldsymbol{X}_{[k]}\boldsymbol{C}_{[k]}\big)^{\top}\big(\boldsymbol{y}-\boldsymbol{X}\tilde{\boldsymbol{\theta}}\big)\Big\Vert_2}$$ $$= -\Big\Vert\big(\boldsymbol{X}_{[k]}\boldsymbol{C}_{[k]}\big)^{\top}\big(\boldsymbol{y}-\boldsymbol{X}\tilde{\boldsymbol{\theta}}\big)\Big\Vert_2.$$

\noindent Therefore, for $\tilde{\boldsymbol{\beta}}_j$ and $\tilde{\boldsymbol{\eta}}_{jj'}$ to be the minimizer's of (\ref{HiGLASSOcriterionfull}), we need $$-\Big\Vert\big(\boldsymbol{X}_{[k]}\boldsymbol{C}_{[k]}\big)^{\top}\big(\boldsymbol{y}-\boldsymbol{X}\tilde{\boldsymbol{\theta}}\big)\Big\Vert_2 + \lambda_1(n) \geq 0,$$

\noindent which implies that

\begin{equation}
    \bigg\Vert\frac{1}{n}\boldsymbol{C}_{[k]}^{\top}\boldsymbol{X}_{[k]}^{\top}\big(\boldsymbol{y}-\boldsymbol{X}\tilde{\boldsymbol{\theta}}\big)\bigg\Vert_2 \leq \frac{\lambda_1(n)}{n}. \label{connect_kkt1}
\end{equation}

\noindent The directional derivative with respect to $\boldsymbol{\eta}_{kk'}$ in the $u$ direction of (\ref{HiGLASSOcriterionfull}) is

$$-\Big(\boldsymbol{X}_{kk'}[\boldsymbol{u} \odot (\boldsymbol{\beta}_k \otimes \boldsymbol{\beta}_{k'})]\Big)^{\top}\big(\boldsymbol{y}-\boldsymbol{X}\boldsymbol{\theta}\big) + \lambda_2(n)$$

\noindent For $\tilde{\boldsymbol{\beta}}_j$ and $\tilde{\boldsymbol{\eta}}_{jj'}$ to be the minimizer's of (\ref{HiGLASSOcriterionfull}), we need $$-\Big(\boldsymbol{X}_{kk'}[\boldsymbol{u} \odot (\tilde{\boldsymbol{\beta}}_k \otimes \tilde{\boldsymbol{\beta}}_{k'})]\Big)^{\top}\big(\boldsymbol{y}-\boldsymbol{X}\tilde{\boldsymbol{\theta}}\big) + \lambda_2(n) \geq 0,$$

\noindent for all $p_kp_{k'}$ dimensional unit vectors $\boldsymbol{u}$. To verify this we must substitute the negative normalized gradient in for $\boldsymbol{u}$, and see when the inequality holds. The negative normalized gradient is given by

$$\boldsymbol{u}^{*} = \frac{\bigg(\boldsymbol{X}_{kk'}\Big[\text{diag}(\tilde{\boldsymbol{\beta}}_k \otimes \tilde{\boldsymbol{\beta}}_{k'})\Big]\bigg)^{\top}\big(\boldsymbol{y}-\boldsymbol{X}\tilde{\boldsymbol{\theta}}\big)}{\Big\Vert\bigg(\boldsymbol{X}_{kk'}\Big[\text{diag}(\tilde{\boldsymbol{\beta}}_k \otimes \tilde{\boldsymbol{\beta}}_{k'})\Big]\bigg)^{\top}\big(\boldsymbol{y}-\boldsymbol{X}\tilde{\boldsymbol{\theta}}\big)\Big\Vert_2}.$$

\noindent Substituting this into our expression for the we get $$-\Big(\boldsymbol{X}_{kk'}[\boldsymbol{u}^{*} \odot (\tilde{\boldsymbol{\beta}}_k \otimes \tilde{\boldsymbol{\beta}}_{k'})]\Big)^{\top}\big(\boldsymbol{y}-\boldsymbol{X}\tilde{\boldsymbol{\theta}}\big)$$

$$= -\frac{\bigg[\bigg(\text{diag}(\tilde{\boldsymbol{\beta}}_k \otimes \tilde{\boldsymbol{\beta}}_{k'})\boldsymbol{X}_{kk'}^{\top}\big(\boldsymbol{y}-\boldsymbol{X}\tilde{\boldsymbol{\theta}}\big)\bigg) \odot (\boldsymbol{\beta}_k \otimes \boldsymbol{\beta}_{k'})\bigg]^{\top}\boldsymbol{X}_{kk'}^{\top}\big(\boldsymbol{y}-\boldsymbol{X}\tilde{\boldsymbol{\theta}}\big)}{\Big\Vert\bigg(\boldsymbol{X}_{kk'}\Big[\text{diag}(\tilde{\boldsymbol{\beta}}_k \otimes \tilde{\boldsymbol{\beta}}_{k'})\Big]\bigg)^{\top}\big(\boldsymbol{y}-\boldsymbol{X}\tilde{\boldsymbol{\theta}}\big)\Big\Vert_2}$$

$$= -\frac{\bigg(\text{diag}(\tilde{\boldsymbol{\beta}}_k \otimes \tilde{\boldsymbol{\beta}}_{k'})\text{diag}(\tilde{\boldsymbol{\beta}}_k \otimes \tilde{\boldsymbol{\beta}}_{k'})\boldsymbol{X}_{kk'}^{\top}\big(\boldsymbol{y}-\boldsymbol{X}\tilde{\boldsymbol{\theta}}\big)\bigg)^{\top}\boldsymbol{X}_{kk'}^{\top}\big(\boldsymbol{y}-\boldsymbol{X}\tilde{\boldsymbol{\theta}}\big)}{\Big\Vert\bigg(\boldsymbol{X}_{kk'}\Big[\text{diag}(\tilde{\boldsymbol{\beta}}_k \otimes \tilde{\boldsymbol{\beta}}_{k'})\Big]\bigg)^{\top}\big(\boldsymbol{y}-\boldsymbol{X}\tilde{\boldsymbol{\theta}}\big)\Big\Vert_2}$$

$$= -\frac{\big(\boldsymbol{y}-\boldsymbol{X}\tilde{\boldsymbol{\theta}}\big)^{\top}\boldsymbol{X}_{kk'}\text{diag}(\tilde{\boldsymbol{\beta}}_k \otimes \tilde{\boldsymbol{\beta}}_{k'})\text{diag}(\tilde{\boldsymbol{\beta}}_k \otimes \tilde{\boldsymbol{\beta}}_{k'})\boldsymbol{X}_{kk'}^{\top}\big(\boldsymbol{y}-\boldsymbol{X}\tilde{\boldsymbol{\theta}}\big)}{\Big\Vert\bigg(\boldsymbol{X}_{kk'}\Big[\text{diag}(\tilde{\boldsymbol{\beta}}_k \otimes \tilde{\boldsymbol{\beta}}_{k'})\Big]\bigg)^{\top}\big(\boldsymbol{y}-\boldsymbol{X}\tilde{\boldsymbol{\theta}}\big)\Big\Vert_2}$$

$$= -\frac{\Big\Vert\bigg(\boldsymbol{X}_{kk'}\Big[\text{diag}(\tilde{\boldsymbol{\beta}}_k \otimes \tilde{\boldsymbol{\beta}}_{k'})\Big]\bigg)^{\top}\big(\boldsymbol{y}-\boldsymbol{X}\tilde{\boldsymbol{\theta}}\big)\Big\Vert_2^2}{\Big\Vert\bigg(\boldsymbol{X}_{kk'}\Big[\text{diag}(\tilde{\boldsymbol{\beta}}_k \otimes \tilde{\boldsymbol{\beta}}_{k'})\Big]\bigg)^{\top}\big(\boldsymbol{y}-\boldsymbol{X}\tilde{\boldsymbol{\theta}}\big)\Big\Vert_2}$$

$$= -\Big\Vert\bigg(\boldsymbol{X}_{kk'}\Big[\text{diag}(\tilde{\boldsymbol{\beta}}_k \otimes \tilde{\boldsymbol{\beta}}_{k'})\Big]\bigg)^{\top}\big(\boldsymbol{y}-\boldsymbol{X}\tilde{\boldsymbol{\theta}}\big)\Big\Vert_2.$$

\noindent Therefore, for $\tilde{\boldsymbol{\beta}}_j$ and $\tilde{\boldsymbol{\eta}}_{jj'}$ to be the minimizer's of (\ref{HiGLASSOcriterionfull}), we need

$$-\Big\Vert\bigg(\boldsymbol{X}_{kk'}\Big[\text{diag}(\tilde{\boldsymbol{\beta}}_k \otimes \tilde{\boldsymbol{\beta}}_{k'})\Big]\bigg)^{\top}\big(\boldsymbol{y}-\boldsymbol{X}\tilde{\boldsymbol{\theta}}\big)\Big\Vert_2 + \lambda_2(n) \geq 0,$$

\noindent which implies that

\begin{equation}
    \bigg\Vert\frac{1}{n}\text{diag}(\tilde{\boldsymbol{\beta}}_k \otimes \tilde{\boldsymbol{\beta}}_{k'})\boldsymbol{X}_{kk'}^{\top}\big(\boldsymbol{y}-\boldsymbol{X}\tilde{\boldsymbol{\theta}}\big)\bigg\Vert_2 \leq \frac{\lambda_2(n)}{n} \label{connect_kkt2}.
\end{equation}

Since (\ref{connect_kkt1}) is equivalent to (\ref{kktmain2}) and (\ref{connect_kkt2}) is equivalent to (\ref{kktint2}), this concludes the proof.

\newpage

\section*{Web Figures}

\newpage

\newpage

\end{document}